\documentclass[acmtocl,acmnow]{acmtrans2m}

\usepackage{amsmath}
\usepackage{amssymb}
\usepackage{graphicx}
\usepackage{xspace,stmaryrd}
\DeclareSymbolFont{letters}{OML}{txmi}{m}{it} 
\usepackage{tikz}
\usetikzlibrary{arrows,automata,shapes,snakes}
\usepackage{color}

\newcommand{\quot}[1]{\ensuremath{#1_{/{\bisim}}}}

\newcommand{\HIDE}[1]{}

\newtheorem{theorem}{Theorem}[section]

\newtheorem{corollary}[theorem]{Corollary}
\newtheorem{proposition}[theorem]{Proposition}
\newtheorem{property}[theorem]{Property}
\newtheorem{lemma}[theorem]{Lemma}
\newdef{definition}[theorem]{Definition}
\newdef{remark}[theorem]{Remark}
\newdef{example}[theorem]{Example}

\newcommand{\Ename}{\mathsf{E}}
\newcommand{\RHSname}{\mathsf{RHS}}
\newcommand{\E}[2]{\Ename^{#1}(#2)}
\newcommand{\RHS}[2]{\RHSname_{#1}(#2)}
\newcommand{\act}{\textit{Act}}

\newcommand{\dom}[1]{\ensuremath{\textsf{dom}(#1)}}

\newcommand{\Nat}{\textit{I\hspace{-.5ex}N}}

\newcounter{soscounter}
\newcommand{\sosrule}[3][]{\refstepcounter{soscounter}(\arabic{soscounter}) #1 \frac{\raisebox{.7ex}{\normalsize{$#2$}}}
                        {\raisebox{-1.0ex}{\normalsize{$#3$}}}}

\newcommand{\bisim}{\;\underline{\hspace*{-0.15ex}
                        \leftrightarrow\hspace*{-0.15ex}}\;}

\def\lparal{\mathbin{\setbox0=\hbox{$\|$}%
        \dimen0=\dp0 \advance\dimen0 -1.5pt \dp0=\dimen0%
        \underline{\kern-1.5pt\box0\kern1.5pt}}}

\newcommand{\ie}{\textit{i.e.}\xspace }
\newcommand{\eg}{\textit{e.g.}\xspace }
\newcommand{\viz}{\textit{viz.}\xspace }

\newcommand{\ibid}{\textit{ibid.}\xspace }

\newcommand{\isdef}{=}
\newcommand{\ap}{{:}}

\renewcommand{\ge}{\mathrel{\geqslant}}

\newcommand{\bnd}[1]{\ensuremath{\mathsf{bnd}(#1)}}
\newcommand{\rhsname}{\ensuremath{\mathsf{rhs}}}
\newcommand{\rhs}[1]{\ensuremath{\rhsname(#1)}}
\newcommand{\occ}[1]{\ensuremath{\mathsf{occ}(#1)}}
\newcommand{\normalise}[1]{\ensuremath{\mathsf{norm}(#1)}}
\newcommand{\srf}[1]{\ensuremath{\mathsf{SRF}(#1)}}

\newcommand{\bool}{\ensuremath{\mathbb{B}}}
\newcommand{\true}{\ensuremath{\mathsf{true}}}
\newcommand{\false}{\ensuremath{\mathsf{false}}}

\newcommand{\mc}[1]{\ensuremath{\mathcal{#1}}}

\newcommand{\rank}[2]{\ensuremath{\mathsf{rank}_{#1}(#2)}}
\newcommand{\block}[2]{\ensuremath{\mathsf{block}_{#1}(#2)}}

\newcommand{\sem}[1]{\ensuremath{[\![ #1 ]\!]}}

\newcommand{\up}{\blacktriangle}
\newcommand{\down}{\blacktriangledown}
\newcommand{\tctx}[2]{\ensuremath{\langle #2, #1 \rangle}}
\newcommand{\fv}[1]{\ensuremath{\nearrow_{#1}}}

\newcommand{\sgbesname}{\ensuremath{\beta}}
\newcommand{\sgbes}[1]{\ensuremath{\sgbesname(#1)}}
\newcommand{\sgformname}{\ensuremath{\varphi}}
\newcommand{\sgform}[1]{\ensuremath{\sgformname(#1)}}
\newcommand{\choice}[1]{\ensuremath{\gamma_{#1}}}
\newcommand{\free}[1]{\ensuremath{\mathsf{free}(#1)}}

\newcommand{\sosref}[1]{\text{(\ref{#1})}}

\title{Structural Analysis of Boolean Equation Systems}

\author{
JEROEN KEIREN and MICHEL A. RENIERS and TIM A.C. WILLEMSE \\
Eindhoven University of Technology}

\begin{abstract}
We analyse the problem of solving Boolean equation systems through the use
of \emph{structure graphs}. The latter are obtained through an elegant
set of Plotkin-style deduction rules.  Our main contribution is that
we show that equation systems with bisimilar structure graphs have the
same solution.  We show that our work conservatively extends earlier work,
conducted by Keiren and Willemse, in which \emph{dependency graphs} were
used to analyse a subclass of Boolean equation systems, \viz, equation
systems in \emph{standard recursive form}. We illustrate our approach by
a small example, demonstrating the effect of simplifying an equation system
through minimisation of its structure graph.

\end{abstract}

\category{}{}{}
\terms{}

\begin{document}

\begin{bottomstuff}
\end{bottomstuff}

\maketitle

\section{Introduction}\label{Sect:intro}

A \emph{Boolean equation system}~\cite{Lar:92,Mad:97} --- equation
system for short --- is a sequence of fixed-point equations, in which
all equations range over the Boolean lattice.  The interest
in equation systems has both practical and theoretical origins.

Equation systems have been used as a uniform framework for solving
traditional verification problems such as the celebrated \emph{model
checking} problem~\cite{Mad:97} and a variety of \emph{behavioural
equivalence checking} problems, see~\cite{Mat:03,CPPW:07}; this has
led to effective tooling, see \eg~\cite{GMLS:07,GMR+:09}.  The size of the
resulting equation system is dependent on the input and the verification
problem: for instance, the global $\mu$-calculus model checking problem
$L \models \phi$, where $L$ is a state space and $\phi$ a formula can
be made to yield equation systems $\E{L}{\phi}$ of size $\mc{O}(|L|
\cdot|\phi|)$, where $|L|$ is the size of the state space and $|\phi|$
the size of the modal formula. As a result, the encoding to equation
systems suffers from a phenomenon akin to the state explosion problem.

From a theoretical stance, the problem of solving an equation
system is intriguing: it is in $\text{NP} \cap \text{co-NP}$, see,
\eg~\cite{Mad:97}. In fact, the problem of solving an equation system
is equivalent to the problem of computing the winner in a \emph{Parity
Game}~\cite{Zie:98}. The latter has been shown to be in $\text{UP} \cap
\text{co-UP}$, see~\cite{Jur:98}. This makes the problem of solving
an equation system a favourable candidate for finding a polynomial
time algorithm, if it exists.  Currently, the algorithm with the best
worst-case time complexity for solving Parity Games, and thereby equation
systems, is the \emph{bigstep} algorithm~\cite{Schewe2007}. This algorithm
has run-time complexity $\mc{O}(n\cdot m^{d/3})$, where $n$ corresponds to
the number of vertices, $m$ the number of edges and $d$ the number of
priorities in the Parity Game  (or equivalently, the number of equations,
the cumulative size of the right-hand sides and the number of fixed-point
sign alternations in an equation system, respectively).

The run-time complexity of the algorithms for solving equation systems
provides a practical motivation for investigating methods for efficiently
reducing the size of equation systems.  In the absence of notions such as
a behaviour of an equation system, an unorthodox strategy in this setting
is the use of bisimulation minimisation techniques.  Nevertheless, recent
work~\cite{KeirenWillemse2009a} demonstrates that such minimisations are
practically cost-effective:  they yield massive reductions of the size
of equation systems, they do not come with memory penalties, and the
time required for solving the original equation system significantly
exceeds the time required for minimisation and subsequent solving of
the minimised equation system.

In \ibid, the minimisations are only obtained for a strict subclass of
equation systems, \viz, equation systems in \emph{standard recursive form
(SRF)}. The minimisation technique relies on a bisimulation minimisation
for a variation of \emph{dependency graphs}~\cite{Mad:97,Kei:06} underlying
the equation systems in SRF.  Such graphs basically reflect the (possibly
mutual) dependencies of the equations in an equation system in SRF.

From a practical viewpoint, the class of equation systems in SRF does
not pose any limitations to the applicability of the method: every
equation system can be brought into SRF without changing the solution
to the proposition variables of the original equation system, and the
transformation comes at the cost of a blow-up in size. Its effects on the
minimising capabilities of bisimulation, however, are unknown, leading
to the first question:
\begin{description}
\item[1] Let $\mc{E}_{/\bisim}$ denote the equation system
$\mc{E}$ minimised with respect to bisimulation and
let $\srf{\mc{E}}$ denote the equation system $\mc{E}$ brought
into SRF. The size of $\mc{E}$ is denoted by $|\mc{E}|$.
Does the following
inequality hold for all $L$ and $\phi$:
\[|\E{L}{\phi}_{/\bisim}| \ge |\srf{\E{L}{\phi}}_{/\bisim}|\]
\end{description}
Furthermore,
it is well-known that the modal $\mu$-calculus is preserved under
bisimulation minimisation of the behavioural state space. However, it is
unknown whether state space minimisation and
minimisation of equation systems encoding a model checking problem
are comparable. This leads to the second question:
\begin{description}
\item[2] 
Let $L_{/\bisim}$ denote the labelled transition system
$L$ minimised with respect to bisimulation.
Does the following inequality hold for all $\phi$:
\[
|\E{L_{/\bisim}}{\phi}| \ge |\E{L}{\phi}_{/\bisim}|
\]
\end{description}

In this paper, we answer both questions positively. In addition, for
both questions we provide examples in which the inequality is in fact
strict. For the second question, our example even illustrates that the
bisimulation reduction of equation
systems can be arbitrarily larger than the reduction of state spaces.

The main problem in obtaining our results is that it is hard to elegantly
capture the structure of an equation system, without resulting in a
parse-tree of the equation system. As a matter of fact, bisimilarity is
required to reflect associativity and commutativity of Boolean operators
such as $\wedge$ and $\vee$ in order to obtain our aforementioned second
result. In addition, the nesting levels of Boolean operators in equation
systems complicate a straightforward definition of bisimilarity for
such general equation systems. We solve these issues by using a set of
deduction rules in Plotkin style~\cite{Plotkin04a} to map the equation
systems onto \emph{structure graphs}. The latter generalise dependency
graphs by dropping the requirement that each vertex necessarily represents
a proposition variable occurring at the left-hand side of some equation
and adding facilities for reasoning about Boolean constants $\true$
and $\false$, and unbound variables.

\paragraph*{Related Work}
This paper extends and improves upon preliminary work presented
in~\cite{ReniersWillemse2009}.

Various types of graphs for equation systems have appeared in the
literature. In~\cite{Mad:97}, Mader considers dependency graphs consisting
of vertices representing equations and edges representing the fact that
one equation depends on the value of another equation. The structure of
the right-hand sides of the equations can in no sense be captured by
these graphs. Kein\"anen~\cite{Kei:06} extends the dependency graphs
of Mader by decorating the vertices with at most one of the Boolean
operators $\wedge$ and $\vee$, and, in addition, a natural number that
abstractly represents the fixed-point sign of the equation.  However,
the dependency graphs of \ibid, only allow for capturing equation
systems in SRF. Keiren and Willemse~\cite{KeirenWillemse2009a} use these
dependency graphs to investigate two notions of bisimulation, \viz,
\emph{strong bisimulation}, and a weakened variation thereof, called
\emph{idempotence-identifying bisimulation}, and their theoretical and
practical use for minimising equation systems.

The dependency graphs of~\cite{Kei:06,KeirenWillemse2009a}, in turn, are
closely related to \emph{Parity Games}~\cite{Zie:98}, in which players
aim to win an infinite game. It has been shown that the latter problem is
equivalent to solving an equation system.  Simulation relations for
Parity Games have been studied in, among others~\cite{FW:06}. Finally,
we mention the framework of \emph{Switching Graphs}~\cite{GP:09}, which
have two kinds of edges: ordinary edges and \emph{switches}, which can be
set to one of two destinations. Switching Graphs are more general than
dependency graphs, but are still inadequate for directly capturing the
structure of the entire class of equation systems. Note that in
this setting, the \emph{$v$-parity loop problem} is equivalent to the
problem of solving Boolean equation systems.

\paragraph*{Outline} For completeness, in
Section~\ref{Sect:preliminaries}, we briefly describe the
formal settings, illustrating the model checking problem and
how this problem can be translated to the problem of solving an
equation system.  Section~\ref{Sect:Structure_Graphs} subsequently
introduces structure graphs and the deduction rules for generating
these from an equation system. Our main results are presented
in Sections~\ref{Sect:normalisation}--\ref{Sect:relation}.
An application of our theory can be found in
Section~\ref{Sect:Application}. Section~\ref{Sect:Conclusions} summarises
our results and outlines future work.

\section{Preliminaries}\label{Sect:preliminaries}

Throughout this section, we assume the existence of two sufficiently
large, disjoint, countable sets of proposition variables $\mc{X}$
and $\tilde{\mc{X}}$.

\subsection{The Modal $\mu$-Calculus}

Labelled transition systems provide a formal, semantical model for the
behaviour of a reactive system. While, in this paper, we are mostly
concerned with Boolean equation systems, our work is motivated by the
model checking problem, \ie, the problem of deciding whether a given
behavioural specification satisfies a temporal or modal formula. For
this reason, we first repeat some basic results from the latter setting
and illustrate its connection to the problem of solving Boolean equation
systems.

\begin{definition} A \emph{labelled transition system} is a three-tuple
$L =\langle S, \act, \to \rangle$, consisting of a finite, non-empty set
of states $S$,  a finite, non-empty set of
actions $\act$ and a transition relation $\to \subseteq S \times \act \times
S$.

\end{definition}
We visualise labelled transition systems by directed, edge-labelled
graphs. In line with this graphical notation, we write $s \xrightarrow{a}
s'$ iff $(s,a,s') \in \to$. The \emph{de facto} behavioural equivalence relation
for labelled transition systems is \emph{strong bisimilarity}, see~\cite{Par:81}.

\begin{definition} Let $L = \langle S, \act, \to \rangle$
be a labelled transition system. A symmetric relation $R \subseteq S \times S$
is a \emph{strong bisimulation} if for all $(s,s') \in R$
\[
\forall{a \in \act, t\in S}:~ s \xrightarrow{a} t
\implies \exists{t' \in S}:~ s' \xrightarrow{a} t' \wedge
(t,t') \in R
\]
States $s \in S$, $s' \in S'$ are \emph{bisimilar} iff there is a
bisimulation relation $R$ that relates states $s$ and $s'$;
\end{definition}

The \emph{propositional modal $\mu$-calculus}, see~\cite{Koz:83}
is a highly-expressive language for analysing behaviours that are
defined through a labelled transition system. We refrain from going
into details, but solely present its grammar and semantics below.
For an accessible contemporary treatment of the modal $\mu$-calculus,
we refer to~\cite{BS:01}.
\begin{definition} Let $\act$ be a finite set of actions.
The set of modal $\mu$-calculus formulae is defined through
the following grammar, which is given directly in positive form:
\[
\phi,\psi ::= \true ~|~ \false ~|~ \tilde X ~|~ \phi \wedge \psi ~|~
\phi \vee \psi ~|~ [A]\phi ~|~ \langle A \rangle \phi ~|~
\nu \tilde X. \phi ~|~ \mu \tilde X. \phi
\]
where $\tilde{X}\in \tilde{\mc{X}}$ is a proposition variable;
$A \subseteq \act$ is a set of actions; $\mu$ is a least fixed
point sign and $\nu$ is a greatest fixed point sign.

\end{definition}
Note that our use of generalised modal operators $[A]\phi$ and
$\langle A \rangle \phi$ is merely for reasons of convenience, and has
no implications for the presented theory in this paper.
Henceforth,  we write $[a]\phi$ instead of $[\{a\}]\phi$
and $[\overline{a}]\phi$ instead of $[\act \setminus \{a\}]\phi$.\\

In a formula $\sigma \tilde X. \phi$, each occurrence of the variable
$\tilde X$ is \emph{bound}. An occurrence of $\tilde X$ in a formula $\phi$ is
bound if it is bound in any subformula of $\phi$.
The set of bound proposition variables in
$\phi$ is denoted $\bnd{\phi}$; the set of proposition variables
that syntactically occur in $\phi$ is denoted $\occ{\phi}$. Formula $\phi$ is said to
be \emph{closed} iff $\occ{\phi} \subseteq \bnd{\phi}$.
We only consider $\mu$-calculus formulae $\phi$ that are \emph{well-formed}, \ie:
\begin{enumerate}
\item there are no two distinct subformulae of $\phi$ that bind
the same proposition variable;
\item for every proposition variable $\tilde X \notin \bnd{\phi}$,
no subformula $\sigma \tilde X. \psi$ occurs in $\phi$.
\end{enumerate}
The well-formedness requirement is a technicality and does not incur a
loss of generality of the theory.\\

Modal $\mu$-calculus formulae $\phi$ are \emph{interpreted} in the
context of a labelled transition system and
an \emph{environment} $\theta : \tilde{\mc{X}} \to 2^S$ that
assigns sets of states to proposition variables. We write
$\theta[\tilde X := S']$ to represent the environment in which
$\tilde X$ receives the value $S'$, and all other proposition variables
have values that coincide with those given by $\theta$.
\begin{definition} Let $L = \langle S, \act, \to \rangle$ be a labelled transition
system and let $\theta : \tilde{\mc{X}} \to 2^S$ be a proposition environment.
The semantics of a $\mu$-calculus formula
$\phi$ is defined inductively as follows:
\[
\begin{array}{lll}
\sem{\true}{\theta} &= & S \\
\sem{\false}{\theta} & = & \emptyset \\
\sem{\tilde X}{\theta} &= & \theta(\tilde X) \\
\sem{\phi \wedge \psi}{\theta} &= & \sem{\phi}{\theta} \cap \sem{\psi}{\theta} \\
\sem{\phi \vee \psi}{\theta} &= & \sem{\phi}{\theta} \cup \sem{\psi}{\theta} \\
\sem{[A]\phi}{\theta} &= & \{s \in S ~|~ \forall s' \in S: \forall a \in A:~
 s \xrightarrow{a} s'
\implies s' \in \sem{\phi}{\theta} \} \\
\sem{\langle A\rangle\phi}{\theta} &= & \{s \in S ~|~ \exists s' \in S: \exists a \in A:~
 s \xrightarrow{a} s'
\wedge s' \in \sem{\phi}{\theta} \} \\
\sem{\nu \tilde X. \phi} &= & \bigcup \{ S' \subseteq S ~|~ S' \subseteq \sem{\phi}{\theta[\tilde X := S']} \} \\
\sem{\mu \tilde X. \phi} &= & \bigcap \{ S' \subseteq S ~|~ \sem{\phi}{\theta[\tilde X := S']} \subseteq S' \}
\end{array}
\]

\end{definition}
The \emph{global} model checking problem, denoted $L,\theta \models
\phi$, is defined as the question whether for all states $s \in S$ of
a given labelled transition system $L = \langle S,\act,\to \rangle$, we
have $s \in \sem{\phi}{\theta}$, for given formula $\phi$ and environment
$\theta$.  The \emph{local} model checking problem, denoted $L, s, \theta
\models \phi$, is the problem whether $s \in \sem{\phi}{\theta}$ for a given
state $s \in S$.  Often,
one is only interested in \emph{closed} formulae: formulae in which no
proposition variable occurs that is not bound by a surrounding fixed
point sign.  Small examples of typical model checking problems can be
found in the remainder of this paper.

\subsection{Boolean Equation Systems}

A Boolean equation system is a finite sequence of least and greatest
fixed point equations, where each right-hand side of an equation is a
proposition formula. For an excellent, in-depth account on Boolean equation systems,
we refer to~\cite{Mad:97}.

\begin{definition}
A \emph{Boolean equation system (BES)} $\mc{E}$ is defined by the following grammar:
\[
\begin{array}{lll}
\mc{E} & ::= & \epsilon ~|~ (\nu X = f)\ \mc{E} ~|~ (\mu X = f)\ \mc{E}\\
f,g    & ::= & \true ~|~ \false ~|~ X ~|~ f \wedge g ~|~ f \vee g 
\end{array}
\]
where  $\epsilon$ is the empty BES; $X \in \mc{X}$ is a proposition variable;
and $f,g$ are proposition formulae. We write $\sigma$ to denote an
arbitrary fixed point sign $\mu$ or $\nu$.
\end{definition}
We only consider equation systems that are \emph{well-formed}, \ie, equation
systems $\mc{E}$, in which a proposition variable $X$ occurs at the left-hand
side in at most a single equation in $\mc{E}$.

In line with the notions of bound and occurring proposition variables
for $\mu$-calculus formulae, we introduce analogue notions for equation
systems.  Let $\mc{E}$ be an arbitrary equation system. The set of
\emph{bound} proposition variables of $\mc{E}$, denoted $\bnd{\mc{E}}$,
is the set of variables occurring at the left-hand side of the equations
in $\mc{E}$. The set of \emph{occurring} proposition variables, denoted
$\occ{\mc{E}}$, is the set of variables occurring at the right-hand
side of some equation in $\mc{E}$.

An equation system $\mc{E}$ is said to be \emph{closed} whenever
$\occ{\mc{E}} \subseteq \bnd{\mc{E}}$.   Intuitively, a (closed) equation
system uniquely assigns truth values to its bound proposition variables,
provided that every bound variable occurs only at the left-hand side of
a single equation in an equation system.  An equation system is said to
be in \emph{simple form} if none of the right-hand sides of the equations that
occur in the equation system contain both $\wedge$- and $\vee$-operators.

Proposition variables occurring in
a proposition formula $f$ are collected in the set $\occ{f}$. The \emph{rank}
of a proposition variable $X \in \bnd{\mc{E}}$,
notation $\rank{\mc{E}}{X}$, is defined as follows:
\[
\begin{array}{l}
\rank{(\sigma Y = f) \mc{E}}{X} = \left \{
\begin{array}{ll}
\rank{\mc{E}}{X} & \text{if $X \not= Y$} \\
\block{\sigma}{\mc{E}} & \text{otherwise}
\end{array}
\right .
\end{array}
\]
where $\block{\sigma}{\mc{E}}$ is defined as:
\[
\begin{array}{l}
\block{\sigma}{\epsilon} =
\left \{
\begin{array}{ll}
0 & \text{if $\sigma = \nu$} \\
1 & \text{otherwise}
\end{array}
\right .
\qquad
\block{\sigma}{(\sigma' Y = f) \mc{E}} = \left \{
\begin{array}{ll}
\block{\sigma}{\mc{E}} & \text{if $\sigma = \sigma'$}\\
1+\block{\sigma'}{\mc{E}} & \text{if $\sigma \not= \sigma'$}\\
\end{array}
\right .
\end{array}
\]
Informally, the rank of a variable $X$ is the $i$-th block of like-signed
equations, containing $X$'s defining equation, counting from
right-to-left and starting at $0$ if the last equation is a greatest
fixed point sign, and $1$ otherwise.\\

Formally, proposition formulae are interpreted in a context of an
\emph{environment} $\eta \ap \mc{X} \to \bool$. For an arbitrary
environment $\eta$, we write $\eta [X:=b]$ for the environment $\eta$
in which the proposition variable $X$ has Boolean value $b$ and all
other proposition variables $X'$ have value $\eta(X')$. The ordering
$\sqsubseteq$ on environments is defined as $\eta \sqsubseteq \eta'$
iff $\eta(X)$ implies $\eta'(X)$ for all $X$.
For reading ease, we do not formally distinguish between a semantic
Boolean value and its representation by $\true$ and $\false$; likewise,
for the operands $\wedge$ and $\vee$.
\begin{definition}\label{def:solution_es}
Let $\eta \ap \mc{X} \to \bool$ be an environment.  The
\emph{interpretation} $\sem{f}{\eta}$ maps a proposition formula $f$
to $\true$ or $\false$:
\begin{align*}
\sem{X}{\eta}      &\isdef \eta(X) \\
\sem{\true}{\eta} &\isdef \true &
\sem{f \wedge g}{\eta} &\isdef \sem{f}{\eta} \wedge \sem{g}{\eta}\\
\sem{false}{\eta} &\isdef \false &
\sem{f \vee g}{\eta} &\isdef \sem{f}{\eta} \vee \sem{g}{\eta}
\end{align*}
The \emph{solution of a BES}, given an environment $\eta$,
is inductively defined as follows:
\[ \begin{array}{lcl}
\sem{\epsilon}{\eta}  & \isdef & \eta \\

\sem{( \sigma X = f )\ \mc{E}}{\eta} & \isdef &
\left \{
\begin{array}{ll}
  \sem{\mc{E}}{ (\eta [X :=  \sem{f}{(\sem{\mc{E}}{\eta[X := \false]})}])}
&  \text{ if $\sigma = \mu$} \\
  \sem{\mc{E}}{ (\eta [X :=  \sem{f}{(\sem{\mc{E}}{\eta[X := \true]})}])}
&  \text{ if $\sigma = \nu$} \\
\end{array}
\right .
\end{array}
\]
\end{definition}
A solution to an equation system verifies every equation, in the sense
that the value at the left-hand side is logically equivalent to the
value at the right-hand side of the equation.  At the same time,
the fixed-point signs of left-most equations \emph{outweigh} the
fixed-point signs of those equations that follow, \ie, the fixed-point
signs of left-most equations are more important. The latter phenomenon
is a result of the nested recursion for evaluating the proposition
$f$ of the left-most equation $(\sigma X = f)$, assuming an extremal
value for $X$.  As a consequence, the solution is order-sensitive: the
solution to $(\mu X = Y)\ (\nu Y = X)$, yielding all $\false$, differs
from the solution to $(\nu Y = X)\ (\mu X = Y)$, yielding all $\true$.
It is exactly this tree-like recursive definition of a solution that makes
it intricately complex.\\

Closed equation systems enjoy the property that the solution to the
equation system is independent of the environment in which it is defined,
\ie, for all environments $\eta,\eta'$, we have $\sem{\mc{E}}{\eta}(X) =
\sem{\mc{E}}{\eta'}(X)$ for all $X \in \bnd{\mc{E}}$.  For this reason,
we henceforth refrain from writing the environment explicitly in all
our considerations dealing with closed equation systems, \ie, we write
$\sem{\mc{E}}$, and $\sem{\mc{E}}(X)$ instead of the more verbose
$\sem{\mc{E}}{\eta}$ and $\sem{\mc{E}}{\eta}(X)$.\\

The following lemma relates the semantics for open equation systems to
that of closed equation systems. We write $\mc{E}[X := b]$, where $X \notin
\bnd{\mc{E}}$ and $b \in \{\true,\false\}$ is a constant, to denote the
equation system in which each syntactic occurrence of $X$ is replaced by
$b$.

\begin{lemma}
\label{lem:substitution}
Let $\mc{E}$ be an equation system, and let $\eta$ be an arbitrary environment.
Assume $X \notin \bnd{\mc{E}}$ is a proposition variable, and let
$b$ be such that $\eta(X) = \sem{b}$.
Then $\sem{\mc{E}}{\eta} = \sem{\mc{E}[X := b]}{\eta}$.

\end{lemma}
\begin{proof}
We show this by induction on the size of $\mc{E}$. The base case for
$\mc{E} = \epsilon$ follows immediately.
As our induction hypothesis, we take
\[
\tag{IH}
\forall \eta,b, X \notin \bnd{\mc{E}}:~
\sem{b} = \eta(X) \implies
\sem{\mc{E}}\eta = \sem{\mc{E}[X := b]}\eta
\]
Assume our induction hypothesis holds
for $\mc{E}$, and let $\eta$ and $b$
be such that $\sem{b} = \eta(X)$. Consider the
equation system $(\nu Y = f)\ \mc{E}$, and assume
$X \notin \bnd{(\nu Y=f)\ \mc{E}}$. Using the semantics of equation systems,
we reason as follows:
\[
\begin{array}{ll}
&
 \sem{(\nu Y = f)\mc{E}}\eta\\
= &
 \sem{\mc{E}}{\eta[Y := \sem{f}{(\sem{\mc{E}}\eta[Y := \true])]}}\\
=^{2 \times IH} &
 \sem{\mc{E}[X:=b]}\eta[Y:=\sem{f}{(\sem{\mc{E}[X := b]}{\eta[Y := \true]})}]\\
=^{\ddagger} &
 \sem{\mc{E}[X:=b]}\eta[Y:=\sem{f[X:=b]}{(\sem{\mc{E}[X:=b]}{\eta[Y:= \true]})}]\\
= &
 \sem{((\nu Y = f)\ \mc{E})\ [X := b]}\eta
\end{array}
\]
where at $\ddagger$, we have used that $\sem{f}{\eta} = \sem{f[X := b]}{\eta}$
for $\sem{b} = \eta(X)$. The case for $(\mu Y = f)\ \mc{E}$ follows the exact
same line of reasoning and is therefore omitted.
\end{proof}

Finally, we introduce some generic shorthand notation.
The operators $\bigsqcap$ and $\bigsqcup$ are used as shorthands for
nested applications of $\land$ and $\lor$. Formally, these are defined as
follows. Let $\lessdot$ be a total
order on $\mc{X} \cup \{\true,\false\}$.  Assuming that $\lessdot$
is lifted to a total ordering on formulae, we define for formula $f$
$\lessdot$-smaller than all formulae in a finite set
$F$:
\begin{align*}
\bigsqcap \emptyset & \isdef \true &
\bigsqcap \{ f \} & \isdef f \wedge f  &
 \bigsqcap (\{f\}\cup F) = f \land \left(\bigsqcap F\right)\\
\bigsqcup \emptyset & \isdef \false &
  \bigsqcup \{ f \} &= f \vee f &
 \bigsqcup (\{f\}\cup F) = f \lor \left(\bigsqcup F\right)
\end{align*}
Let $X = f$ be a non-fixed point equation, where
$f$ is a proposition formula and $X$ is a proposition variable. Assuming that
$X$ is $\lessdot$-smaller than all left-hand side variables in the equations
in a finite set of non-fixed point equations $E$, we define:
\begin{align*}
\sigma \{X = f\} & \isdef (\sigma X = f)
&
\sigma( \{X = f \} \cup E) &\isdef (\sigma X = f) \sigma E
\end{align*}
Note the non-standard duplication of formulae in case the operators
$\bigsqcap$ and $\bigsqcup$ are applied to singleton sets. While this
has no semantic impact, the reasons for the duplication of the least formula
will become apparent in the next section.

\subsection{Boolean Equation Systems for Model Checking}

An obvious strategy for solving a typical model checking problem is through
the use of Tarski's approximation schemes for computing the solution
to the fixed points of monotone operators in a complete lattice, see
\eg~\cite{Tar:55}. More advanced techniques employ intermediate
formalisms such as Boolean equation systems for solving the verification
problem.

Below, we provide the translation of the model checking problem to the
problem of solving a Boolean equation system.
The transformer $\Ename$ reduces the global model checking problem
$L, \eta \models \phi$ to the problem of solving an equation system.
\begin{definition}
\label{def:transformation} Assume $L = \langle S,\act,\to \rangle$
is a labelled transition system. Let $\phi$ be an arbitrary modal
$\mu$-calculus formula over $\act$. Suppose that for every proposition
variable $\tilde X \in \occ{\phi} \cup \bnd{\phi}$, we have a set of
fresh proposition variables $\{X_s ~|~ s \in S \} \subseteq \mc{X}$.
\[
\begin{array}{lll}
\E{L}{b} &=& \epsilon \\
\E{L}{X} &=& \epsilon \\
\E{L}{f \wedge g} &=& \E{L}{f}\ \E{L}{g} \\
\E{L}{f \vee g} &=& \E{L}{f}\ \E{L}{g} \\
\E{L}{[A]f} & =& \E{L}{f} \\
\E{L}{\langle A\rangle f} & =& \E{L}{f} \\
\E{L}{\sigma X.~ f} &=&
(\sigma  \{ (X_s = \RHS{s}{f} ) ~|~
s \in S\} )\ \E{L}{f}\\
\\
\RHS{s}{b} &=& b \\
\RHS{s}{X} &=& X_s \\
\RHS{s}{f \wedge g} &=& \RHS{s}{f} \wedge \RHS{s}{g}\\
\RHS{s}{f \vee g} &=& \RHS{s}{f} \vee \RHS{s}{g} \\
\RHS{s}{[A]f} & =&  \bigsqcap \{\RHS{t}{f} ~|~ a \in A, s \xrightarrow{a} t\} \\
\RHS{s}{\langle A\rangle f} & =&  \bigsqcup \{\RHS{t}{f} ~|~ a \in A, s \xrightarrow{a} t\}\\
\RHS{s}{\sigma X.~ f} &=& X_s
\end{array}
\]

\end{definition}
Observe that the definition of $\Ename$ provided here coincides
(semantically) with the definition given in~\cite{Mad:97} for modal
$\mu$-calculus formulae $\phi$;
the only deviation is a syntactic one, ensuring that the $[\_]$ and
$\langle\_\rangle$ modalities are mapped onto proposition formulae with
$\wedge$, and $\vee$ as their main logical connectives in case there is
a non-empty set of emanating transitions.

The relation between the original local model checking problem and the problem
of solving a Boolean equation system is stated by the theorem below.
\begin{theorem}[\cite{Mad:97}]
Assume $L = \langle S, \act, \to \rangle$ is a labelled transition
system. Let $\sigma \tilde X. f$ be an arbitrary modal $\mu$-calculus formula,
and let $\theta$ be an arbitrary environment. Then:
\[
L, s, \theta \models \sigma \tilde X. f
\text{ iff }
(\sem{\E{L}{\sigma \tilde X.f}}{(\lambda Y_t \in \mc{X}.~
t \in \theta(\tilde Y))})(X_s)
\]

\end{theorem}
The example below illustrates the above translation and theorem.
\begin{example}
Consider the labelled transition system (depicted below), modelling
mutual exclusion between two readers and a single writer.

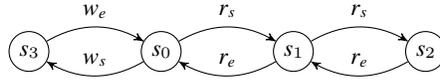
\begin{center}
\begin{tikzpicture}[->,>=stealth',node distance=50pt]
\tikzstyle{every state}=[minimum size=15pt, inner sep=2pt, shape=circle]

\node [state] (naught) {\small $s_0$};
\node [state] (one) [right of=naught] {\small $s_1$};
\node [state] (two) [right of=one] {\small $s_2$};
\node [state] (three) [left of=naught] {\small $s_3$};

\draw
  (naught) edge[bend left] node [above] {\small $r_s$} (one)
  (one) edge[bend left] node [above]    {\small $r_s$} (two)
  (two) edge[bend left] node [above]    {\small $r_e$} (one)
  (one) edge[bend left] node [above]    {\small $r_e$} (naught)
  (naught) edge[bend left] node [above] {\small $w_s$} (three)
  (three) edge[bend left] node [above]  {\small $w_e$} (naught);
\end{tikzpicture}
\end{center}

\noindent
Reading is started using an action $r_s$ and action $r_e$ indicates
its termination. Likewise for writing.  The verification problem $\nu
\tilde X. \mu \tilde Y. ~\langle r_s\rangle \tilde X \vee \langle \overline{r_s}
\rangle \tilde Y$, modelling that on some path, a reader
can infinitely often start reading, translates to the following equation
system using the translation $\Ename$:

$$
\small
\begin{array}{l}
(\nu X_{s_0} = Y_{s_0})\
(\nu X_{s_1} = Y_{s_1})\
(\nu X_{s_2} = Y_{s_2})\
(\nu X_{s_3} = Y_{s_3})\\
(\mu Y_{s_0} = (X_{s_1} \vee X_{s_1}) \vee (Y_{s_3} \vee Y_{s_3}))\\
(\mu Y_{s_1} = (X_{s_2} \vee X_{s_2}) \vee (Y_{s_0} \vee Y_{s_0}))\\
(\mu Y_{s_2} = \false \vee (Y_{s_1} \vee Y_{s_1}))\\
(\mu Y_{s_3} = \false \vee (Y_{s_0} \vee Y_{s_0}))
\end{array}
$$
\noindent
Observe that, like the original $\mu$-calculus formula, which has mutual
dependencies between $\tilde X$ and $\tilde Y$, the resulting equation
system has mutual dependencies between the indexed $X$ and $Y$
variables.
Solving the resulting equation system leads to $\true$ for all bound
variables; $X_{s_i} = \true$, for arbitrary state $s_i$,
implies that the property holds in state $s_i$. Furthermore,
note that the right-hand sides of the resulting equation system
can be rewritten using standard rules of logic, removing, \eg, all
occurrences of $\false$.
\end{example}

\section{Structure Graphs for Boolean Equation Systems}
\label{Sect:Structure_Graphs}

A large part of the complexity of equation systems is attributed to the
mutual dependencies between the equations. These intricate dependencies
are neatly captured by \emph{structure graph}s. Another issue is how
to deal with variables that are not defined in the equation system
but are used in proposition formulae.  We first introduce structure
graphs, and define the well-known notion of bisimilarity on those.
In Section \ref{Subsect:SOSforBES}, we define how a structure graph can
be obtained from a formula in the context of an equation system. In
Section \ref{Subsect:SGtoBES}, we define how an equation system can
be associated to a structure graph assuming that it satisfies some
well-formedness constraints.


\label{subsec:sg}

\begin{definition}
Given a set of proposition variables $\mc{X}$.
A structure graph over $\mc{X}$ is a vertex-labelled graph
$\mc{G} = \langle T, t, \to, d, r, \nearrow \rangle$, where:

\begin{itemize}
\item $T$ is a finite set of vertices;
\item $t \in T$ is the initial vertex;
\item $\to \subseteq T \times T$ is a dependency relation;
\item $d\ap T \mapsto \{ \up, \down, \top, \perp \}$ is a vertex decoration mapping;
\item $r\ap T \mapsto \Nat$ is a vertex ranking mapping;
\item $\nearrow \ap T \mapsto \mc{X}$ is a free variable mapping.
\end{itemize}
\end{definition}

A structure graph allows for capturing the dependencies between bound
variables and (sub)formulae occurring in the equations of such bound
variables.  Intuitively, the decoration mapping $d$ reflects whether the top symbol of a proposition formula is $\true$ (represented by $\top$), $\false$
(represented by $\perp$), a conjunction (represented by $\up$), or a
disjunction (represented by $\down$). The vertex ranking mapping $r$ indicates the rank of a vertex.
The free variable mapping indicates whether a vertex represents a free variable.
Note that each vertex can have
at most one rank, at most one decoration $\star \in \{ \up, \down,
\top, \perp \}$, and at most one free variable $\fv{X}$. We sometimes write
$t$ to refer to a structure graph $\langle T, t, \to, d, r, \nearrow \rangle$,
where $t$ is in fact the root of the structure graph.
One can easily define bisimilarity on structure graphs.

\begin{definition}
Let $\mc{G} = \langle T, t, \to, d, r, \nearrow \rangle$ and $\mc{G}' = \langle T', t', \to', d', r', \nearrow' \rangle$
be structure graphs. A relation $R \subseteq T \times T'$ is a bisimulation relation if for all $(u,u') \in R$
\begin{itemize}
\item $d(u) = d'(u')$, $r(u) = r'(u')$, and $\nearrow(u) = \nearrow'(u')$;
\item for all $v \in T$, if $u \rightarrow v$, then $u' \rightarrow' v'$ for some $v' \in T'$ such that $(v,v') \in R$;
    \item for all $v' \in T'$, if $u' \rightarrow' v'$, then $u \rightarrow v$ for some $v \in T$ such that $(v,v') \in R$.
\end{itemize}
Two vertices $u$ and $u'$ are bisimilar, notation $u \bisim u'$ if there exists a bisimulation relation $R$ such that $(u,u')\in R$.
\end{definition}

\subsection{Structured Operational Semantics for equation systems}
\label{Subsect:SOSforBES}

Next, we define structure graphs for arbitrary equation systems $\mc{E}$ and proposition formulae $t$. We use Plotkin-style Structural Operational Semantics \cite{Plotkin04a} to associate a structure graph with a formula $f$ in the context of a equation system $\mc{E}$, notation $\tctx{\mc{E}}{f}$. The deduction rules define a relation $\_\rightarrow\_$ and predicates $\_ \pitchfork n$ (for $n \in \Nat$), $\_\nearrow X$ (for $X \in \mc{X}$), $\_\top$, $\_\perp$, $\_\up$, and $\_\down$.
In the deduction rules also negative premises are used, see \cite{Mousavi05-IC} for an overview.

The notations used in the deduction rules are slightly different from those used in the structure graphs. The predicate $t \nearrow X$ represents $\nearrow(t) = X$, the predicate $t \pitchfork n$ represents $r(t) = n$, for $\star \in \{ \up,\down,\top,\bot \}$, $t\star$ represents $d(t) = \star$.
The notation $t \not\pitchfork$ represents $\neg (t \pitchfork n)$ for all $n \in \Nat$.

First, as we are dealing with possibly open equation systems, free
variables are labelled as such:

$$
\sosrule[\label{sos:fv_ax}]{X \notin \bnd{\mc{E}}}{\tctx{\mc{E}}{X} \fv{X}}
$$
In addition, vertices representing bound proposition variables are labelled by a
natural number representing the rank of the variable in the equation system:

\[
\sosrule[\label{sos:rank_ax}]{X \in \bnd{\mc{E}}}{\tctx{\mc{E}}{X} \pitchfork \rank{\mc{E}}{X}}
\]


Note that this deduction rules will not allow the derivation of a rank for a proposition variable that is not bound by the equation system.

In Boolean equation systems, conjunction and
disjunction are binary operators. A question that needs
to be answered is ``How to capture this structure in the structure
graph?'' One way of doing so would be to precisely reflect the structure
of the proposition formula. For a formula of the form $X \land (Y
\land Z)$ in the context of an empty equation system this results in the first structure graph depicted below:

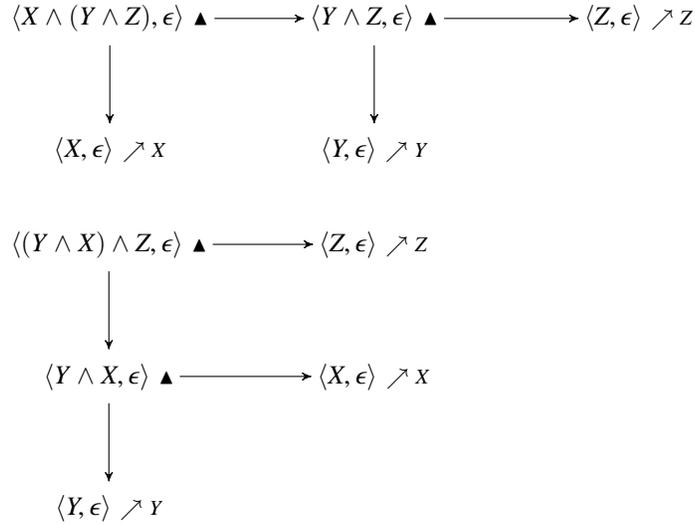
\begin{tabular}{l}
\begin{tikzpicture}[->,>=stealth',node distance=50pt]
\tikzstyle{every state}=[draw=none,minimum size=20pt, inner sep=2pt, shape=rectangle]

\node[state] (f) {$\tctx{\epsilon}{X \land (Y \land Z)}\ \text{\footnotesize{$\up$}}$};
\node[state] (fake) [right of=f] {};
\node[state] (X) [below of=f] {$\tctx{\epsilon}{X}\ \text{\footnotesize{$\nearrow X$}}$};
\node[state] (sf) [right of=fake] {$\tctx{\epsilon}{Y \land Z}\ \text{\footnotesize{$\up$}}$};
\node[state] (Y) [below of=sf] {$\tctx{\epsilon}{Y}\ \text{\footnotesize{$\nearrow Y$}}$};
\node[state] (fake2) [right of=sf] {};
\node[state] (Z) [right of=fake2] {$\tctx{\epsilon}{Z}\ \text{\footnotesize{$\nearrow Z$}}$};

\draw (f) edge (X)
      (f) edge (sf)
      (sf) edge (Y)
      (sf) edge (Z);
\end{tikzpicture}
\\
\\
\begin{tikzpicture}[->,>=stealth',node distance=50pt]
\tikzstyle{every state}=[draw=none,minimum size=20pt, inner sep=2pt, shape=rectangle]

\node[state] (f) {$\tctx{\epsilon}{(Y \land X) \land Z}\ \text{\footnotesize{$\up$}}$};
\node[state] (fake) [right of=f] {};
\node[state] (XY) [below of=f] {$\tctx{\epsilon}{Y\land X}\ \text{\footnotesize{$\up$}}$};
\node[state] (Y) [below of=XY] {$\tctx{\epsilon}{Y}\ \text{\footnotesize{$\nearrow Y$}}$};
\node[state] (Z) [right of=fake] {$\tctx{\epsilon}{Z}\ \text{\footnotesize{$\nearrow Z$}}$};
\node[state] (X) [below of=Z] {$\tctx{\epsilon}{X}\ \text{\footnotesize{$\nearrow X$}}$};

\draw (f) edge (XY)
      (f) edge (Z)
      (XY) edge (Y)
      (XY) edge (X);
\end{tikzpicture}
\end{tabular}

A drawback of this solution is that, in general, the logical equivalence
between $X \land (Y \land Z)$ and $(Y \land X) \land Z$ (see the second structure graph above)
is not reflected by bisimilarity. Retaining this logical equivalence
(and hence associativity and commutativity) of both conjunction and
disjunction is desirable, and, in fact, one of our major goals.

The logical connectives for conjunction
($\land$) and disjunction ($\lor$) may occur nested in a formula.
This is solved by reflecting a change
in leading operator in the structure graph. So the anticipated structure
of the structure graph for $X \land (Y \land (Z \lor X))$, where, again,
we assume that the equation system contains no equations, is:

\begin{tikzpicture}[->,>=stealth',node distance=50pt]
\tikzstyle{every state}=[draw=none,minimum size=20pt, inner sep=2pt, shape=rectangle]

\node[state] (f) {$\tctx{\epsilon}{X \land (Y \land (Z \lor X))}\ \text{\footnotesize{$\up$}}$};
\node[state] (fake) [right of=f] {};
\node[state] (sf) [right of=fake] {$\tctx{\epsilon}{Z \lor X}\ \text{\footnotesize{$\down$}}$};
\node[state] (fake2) [right of=sf] {};
\node[state] (Z) [right of=fake2] {$\tctx{\epsilon}{Z}\ \text{\footnotesize{$\nearrow Z$}}$};
\node[state] (Y) [below of=f] {$\tctx{\epsilon}{Y}\ \text{\footnotesize{$\nearrow Y$}}$};
\node[state] (X) [below of=sf] {$\tctx{\epsilon}{X}\ \text{\footnotesize{$\nearrow X$}}$};

\draw (f) edge (X)
      (f) edge (Y)
      (f) edge (sf)
      (sf) edge (Z) edge (X);
\end{tikzpicture}

This can be elegantly achieved by means of the following deduction rules for the decorations and the dependency transition relation $\rightarrow$:
$$
\sosrule[\label{sos:true_ax}]{}{\tctx{\mc{E}}{\true} \top}
\qquad
\sosrule[\label{sos:false_ax}]{}{\tctx{\mc{E}}{\false} \perp}
\qquad
\sosrule[\label{sos:and_ax}]{}{\tctx{\mc{E}}{f \land f'} \up}
\qquad
\sosrule[\label{sos:or_ax}]{}{\tctx{\mc{E}}{f \lor f'}\down}
$$
\newline
$$
\sosrule[\label{sos:and_left_not_ranked}]{\tctx{\mc{E}}{f}\up \quad \tctx{\mc{E}}{f} \not\pitchfork \quad
\tctx{\mc{E}}{f} \rightarrow \tctx{\mc{E}}{g}}{\tctx{\mc{E}}{f \land f'} \rightarrow \tctx{\mc{E}}{g}}
\qquad
\sosrule[\label{sos:and_right_not_ranked}]{\tctx{\mc{E}}{f'}\up \quad \tctx{\mc{E}}{f'} \not\pitchfork \quad \tctx{\mc{E}}{f'} \rightarrow \tctx{\mc{E}}{g'}}{\tctx{\mc{E}}{f \land f'} \rightarrow \tctx{\mc{E}}{g'}}
$$
$$
\sosrule[\label{sos:or_left_not_ranked}]{\tctx{\mc{E}}{f}\down \quad \tctx{\mc{E}}{f} \not\pitchfork \quad \tctx{\mc{E}}{f} \rightarrow \tctx{\mc{E}}{g}}{\tctx{\mc{E}}{f \lor f'} \rightarrow \tctx{\mc{E}}{g}}
\qquad
\sosrule[\label{sos:or_right_not_ranked}]{\tctx{\mc{E}}{f'}\down \quad \tctx{\mc{E}}{f} \not\pitchfork \quad \tctx{\mc{E}}{f'} \rightarrow \tctx{\mc{E}}{g'}}{\tctx{\mc{E}}{f \lor f'} \rightarrow \tctx{\mc{E}}{g'}}
$$
\newline
$$
\sosrule[\label{sos:and_not_and_left}]{\neg \tctx{\mc{E}}{f} \up}{\tctx{\mc{E}}{f \land f'} \rightarrow \tctx{\mc{E}}{f}}
\qquad
\sosrule[\label{sos:and_not_and_right}]{\neg \tctx{\mc{E}}{f'} \up}{\tctx{\mc{E}}{f \land f'} \rightarrow \tctx{\mc{E}}{f'}}
$$
$$
\sosrule[\label{sos:or_not_or_left}]{\neg \tctx{\mc{E}}{f} \down}{\tctx{\mc{E}}{f \lor f'} \rightarrow \tctx{\mc{E}}{f}}
\qquad
\sosrule[\label{sos:or_not_or_right}]{\neg \tctx{\mc{E}}{f'} \down}{\tctx{\mc{E}}{f \lor f'} \rightarrow \tctx{\mc{E}}{f'}}
$$

$$
\sosrule[\label{sos:and_ranked_left}]{\tctx{\mc{E}}{f} \pitchfork n}{\tctx{\mc{E}}{f \land f'} \rightarrow \tctx{\mc{E}}{f}}
\qquad
\sosrule[\label{sos:and_ranked_right}]{\tctx{\mc{E}}{f'} \pitchfork n}{\tctx{\mc{E}}{f \land f'} \rightarrow \tctx{\mc{E}}{f'}}
$$
$$
\sosrule[\label{sos:or_ranked_left}]{\tctx{\mc{E}}{f} \pitchfork n}{\tctx{\mc{E}}{f \lor f'} \rightarrow \tctx{\mc{E}}{f}}
\qquad
\sosrule[\label{sos:or_ranked_right}]{\tctx{\mc{E}}{f'} \pitchfork n}{\tctx{\mc{E}}{f \lor f'} \rightarrow \tctx{\mc{E}}{f'}}
$$
Rules (\ref{sos:true_ax}-\ref{sos:or_ax}) describe the axioms for decoration.
The first four deduction rules (\ref{sos:and_left_not_ranked}-\ref{sos:or_right_not_ranked}) for $\rightarrow$ are introduced to flatten the nesting hierarchy of the same connective. They can be used to deduce that $X \land (Y \land Z) \rightarrow Y$. Deduction rules \ref{sos:and_not_and_left}-\ref{sos:or_ranked_right} describe the dependencies in case there is no flattening possible anymore (by absence of structure). The deduction rules \ref{sos:and_not_and_left}-\ref{sos:or_not_or_right} deal with the case that a subformula has no $\up$ or $\down$. Deduction rules \ref{sos:and_left_not_ranked}-\ref{sos:or_right_not_ranked} works for the situation that the subformula has a $\up$ or $\down$ but that this is not caused by a recursion variable. The deduction rules \ref{sos:and_ranked_left}-\ref{sos:or_ranked_right} deal with the case that the subformula represents a bound variable.


%
%
%

Finally, we present deduction rules that describe how the structure of a vertex representing a variable is derived from the right-hand side of the corresponding equation. Observe that the deduction rules only have to deal with the case that a defining equation for the recursion variable $X$ has been found in the Boolean equation system.
Deduction rules \ref{sos:var_edge_simple_rhs} and \ref{sos:var_edge_ranked_rhs} define the dependency relation for the case that the right-hand side is a variable or a constant. Deduction rules \ref{sos:var_edge_and_non_ranked_rhs} and \ref{sos:var_edge_or_non_ranked_rhs} do this for the cases it is a proposition formula that is not a variable or a constant.

$$
 \sosrule[\label{sos:var_and}]{\sigma X =f \in \mc{E} \quad \tctx{\mc{E}}{f} \up \quad \tctx{\mc{E}}{f}\not\pitchfork}
 {\tctx{\mc{E}}{X} \up}
 \qquad
 \sosrule[\label{sos:var_or}]{\sigma X =f \in \mc{E} \quad \tctx{\mc{E}}{f} \down \quad \tctx{\mc{E}}{f}\not\pitchfork}
 {\tctx{\mc{E}}{X} \down}
$$
 $$
\sosrule[\label{sos:var_edge_simple_rhs}]{\sigma X =f \in \mc{E} \quad \neg \tctx{\mc{E}}{f} \down \quad \neg \tctx{\mc{E}}{f} \up}{\tctx{\mc{E}}{X} \rightarrow \tctx{\mc{E}}{f}}
\qquad
\sosrule[\label{sos:var_edge_ranked_rhs}]{\sigma X = f\in\mc{E} \quad \tctx{\mc{E}}{f} \pitchfork n}{\tctx{\mc{E}}{X} \rightarrow \tctx{\mc{E}}{f}}
$$
$$
\sosrule[\label{sos:var_edge_and_non_ranked_rhs}]{\sigma X = f\in\mc{E} \quad \tctx{\mc{E}}{f} \rightarrow \tctx{\mc{E}}{g}
\quad \tctx{\mc{E}}{f} \up \quad \tctx{\mc{E}}{f}\not\pitchfork}{\tctx{\mc{E}}{X} \rightarrow \tctx{\mc{E}}{g}}
$$
$$
\sosrule[\label{sos:var_edge_or_non_ranked_rhs}]{\sigma X = f\in\mc{E} \quad \tctx{\mc{E}}{f} \rightarrow \tctx{\mc{E}}{g}
\quad \tctx{\mc{E}}{f} \down \quad \tctx{\mc{E}}{f}\not\pitchfork}{\tctx{\mc{E}}{X} \rightarrow \tctx{\mc{E}}{g}}
$$

\begin{example}\label{ex:structure_graph}  An equation system $\mc{E}$ (see left)
and its associated structure graph (see right). Observe that the term $X
\wedge Y$ is shared by the equations for $X$ and $Y$, and appears only
once in the structure graph as an unranked vertex. There is no equation
for $Z$; this is represented by term $Z$, decorated only by the label $\nearrow Z$.
The subterm $Z \vee W$ in the equation for $W$
does not appear as a separate vertex in the structure graph, since the
disjunctive subterm occurs within the scope of another disjunction.

\noindent
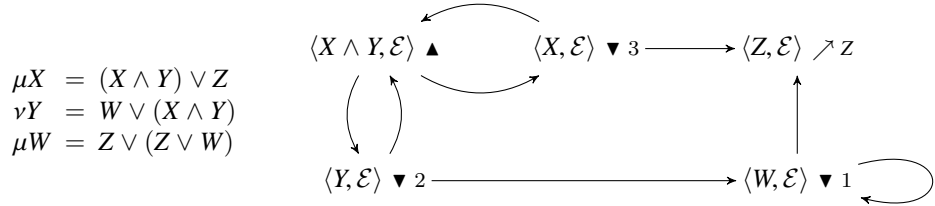
\parbox{.3\textwidth}
{
\[
\begin{array}{lcl}
\mu X &=& (X \wedge Y) \vee Z\\
\nu Y &=& W \vee (X \wedge Y)\\
\mu W &=& Z \vee (Z \vee W)
\end{array}
\]
}
\quad
\parbox{.4\textwidth}
{
\begin{tikzpicture}[->,>=stealth',node distance=50pt]
\tikzstyle{every state}=[shape=rectangle,draw=none,minimum width=30pt, minimum height=20pt, inner sep=2pt]

\node[state] (X) {$\tctx{\mc{E}}{X}\ \text{\footnotesize{$\down\ 3$}}$};
\node[state] (Z) [right of=X,xshift=30pt] {$\tctx{\mc{E}}{Z}\ \text{\footnotesize{$\nearrow Z$}}$};
\node[state] (XY) [left of=X,xshift=-30pt] {$\tctx{\mc{E}}{X \wedge Y}\ \text{\footnotesize{$\up$}}$};
\node[state] (Y) [below of=XY] {$\tctx{\mc{E}}{Y}\ \text{\footnotesize{$\down\ 2$}}$};
\node[state] (W) [below of=Z] {$\tctx{\mc{E}}{W}\ \text{\footnotesize{$\down\ 1$}}$};

\draw (X) edge  (Z) edge[bend right] (XY)
      (Y) edge (W) edge [bend right] (XY)
      (XY) edge [bend right] (Y) edge [bend right] (X)
      (W) edge (Z) edge [loop right] (W);
\end{tikzpicture}
}
\end{example}

Given a formula $f$ and an equation system $\mc{E}$, $\tctx{\mc{E}}{f}$ denotes the part of the structure graph generated by the deduction rules that is reachable from the vertex $\tctx{\mc{E}}{f}$.

%


\begin{lemma}\label{congr}
Let $\mc{E}$ be an equation system. Let $f$, $f'$, $g$ and $g'$ be arbitrary proposition formulae such that $\tctx{\mc{E}}{f} \bisim \tctx{\mc{E}}{f'}$ and $\tctx{\mc{E}}{g} \bisim \tctx{\mc{E}}{g'}$. Then the following hold:
\[ \tctx{\mc{E}}{f \land g} \bisim \tctx{\mc{E}}{f' \land g'},
\quad \tctx{\mc{E}}{f \lor g} \bisim \tctx{\mc{E}}{f' \lor g'}
\]
\end{lemma}

\begin{proof}
Suppose that bisimilarity of $\tctx{\mc{E}}{f}$ and $\tctx{\mc{E}}{f'}$ is witnessed by $R$ and the bisimilarity of $\tctx{\mc{E}}{g}$ and $\tctx{\mc{E}}{g'}$ is witnessed by $S$. The relation $\{ (\tctx{\mc{E}}{f \land g},\tctx{\mc{E}}{f' \land g'}) \} \cup R \cup S$ is a bisimulation relation that proves bisimilarity of $\tctx{\mc{E}}{f\land g}$ and $\tctx{\mc{E}}{f'\land g'}$. Similarly, $\{ (\tctx{\mc{E}}{f \lor g},\tctx{\mc{E}}{f' \lor g'}) \} \cup R \cup S$ is a bisimulation relation that proves bisimilarity of $\tctx{\mc{E}}{f\lor g}$ and $\tctx{\mc{E}}{f'\lor g'}$.
\end{proof}

The following lemma indicates that we achieved the goal that bisimilarity
on structure graphs respects logical equivalences such as commutativity,
associativity and a weak form of idempotence for the $\wedge$ and
$\vee$ operators.

\begin{lemma}\label{props}
Let $\mc{E}$ be an equation system. Let $f$, $f'$, and $f''$ be arbitrary proposition formulae. Then the following hold:
\[
\begin{array}{rcl}
\tctx{\mc{E}}{(f \land f') \land f''} &\bisim& \tctx{\mc{E}}{f \land (f' \land f'')},\\
\tctx{\mc{E}}{(f \lor f') \lor f''} &\bisim& \tctx{\mc{E}}{f \lor (f' \lor f'')}, \\
\tctx{\mc{E}}{f \land f'} &\bisim& \tctx{\mc{E}}{f' \land f}, \\
\tctx{\mc{E}}{f \lor f'} &\bisim& \tctx{\mc{E}}{f' \lor f}, \\
\tctx{\mc{E}}{(f \land f) \land f'} &\bisim& \tctx{\mc{E}}{f \land f'}, \\
\tctx{\mc{E}}{(f \lor f) \lor f'} &\bisim& \tctx{\mc{E}}{f \lor f'}
\end{array}
\]
\end{lemma}

\begin{proof}
The proofs are easy. For example, the bisimulation relation that witnesses
bisimilarity of $\tctx{\mc{E}}{(f \land f') \land f''}$ and $\tctx{\mc{E}}{f \land (f' \land f'')}$
is the relation that relates all formulae of the form $\tctx{\mc{E}}{(g \land g') \land g''}$ and $\tctx{\mc{E}}{g \land (g' \land g'')}$ and additionally contains the
identity relation on structure graphs. Proofs of the ``transfer conditions''
are easy as well. As an example, suppose that $\tctx{\mc{E}}{(g \land g') \land g''}
\rightarrow \tctx{\mc{E}}{h}$ for some formula $h$. In case this transition is due to
$\tctx{\mc{E}}{g \land g'} \up$ and $\tctx{\mc{E}}{g \land g'} \rightarrow \tctx{\mc{E}}{h}$, one of the cases that occurs for $\tctx{\mc{E}}{g
\land g'} \rightarrow \tctx{\mc{E}}{h}$ is that $\tctx{\mc{E}}{g} \up$ and $\tctx{\mc{E}}{g} \rightarrow \tctx{\mc{E}}{h}$. We obtain
$\tctx{\mc{E}}{g \land (g' \land g'')} \rightarrow \tctx{\mc{E}}{h}$. Since $\tctx{\mc{E}}{h}$ and $\tctx{\mc{E}}{h}$ are related,
this finishes the proof of the transfer condition in this case. All
other cases are similar or at least equally easy.
\end{proof}

\begin{corollary}
\label{cor:capcup_props}
Let $\mc{E}$ be an equation system.
Let $F$ and $G$ be arbitrary finite sets of proposition formulae such that (1) for all $f\in F$ there exists $g \in G$ with $\tctx{\mc{E}}{f} \bisim \tctx{\mc{E}}{g}$, and, vice versa, (2) for all $g \in G$ there exists $f \in F$ with $\tctx{\mc{E}}{g} \bisim \tctx{\mc{E}}{f}$. Then, $\tctx{\mc{E}}{\bigsqcap F} \bisim \tctx{\mc{E}}{\bigsqcap G}$ and $\tctx{\mc{E}}{\bigsqcup F} \bisim \tctx{\mc{E}}{\bigsqcup G}$.
\end{corollary}

\begin{proof}
The corollary follows immediately from the congruence of $\land$ and $\lor$ (Lemma \ref{congr}) and commutativity and associativity of those (Lemma \ref{props}).
\end{proof}

Idempotence of $\land$ and $\lor$, and more involved logical equivalences
such as distribution and absorption are not captured by isomorphism or
even bisimilarity on the structure graphs. The reason is that, for an arbitrary equation system $\mc{E}$ and variable $X$, the vertex associated with $\tctx{\mc{E}}{X \land X}$ will be decorated by $\up$, whereas the vertex associated with $\tctx{\mc{E}}{X}$ is not!

\subsection{Translating Structure Graphs to Equation Systems}\label{Subsect:SGtoBES}

Next, we show how, under some mild conditions, a formula and equation
system can be obtained from a structure graph. Later in the paper this
transformation will be used and proved correct.

A structure graph $\mc{G} = \langle T,t,\to,d,r\nearrow \rangle$ is called \emph{BESsy}~if it satisfies the following constraints:
\begin{itemize}
\item a vertex $t$ decorated by $\top, \perp$ or $\fv{X}$ for some $X$ has no successor w.r.t.\ $\to$.
\item a vertex is decorated by $\up$ or $\down$ or a rank iff it has a successor w.r.t.\ $\to$.
\item a vertex with multiple successors w.r.t.\ $\to$, is decorated with $\up$ or $\down$.
\item every cycle contains a vertex with a rank.
\end{itemize}
Observe that BESsyness is preserved under bisimilarity:

\begin{lemma}
Let $\mc{G}$ and $\mc{G}'$ be bisimilar structure graphs.  Then, $\mc{G}$
is BESsy if, and only if, $\mc{G}'$ is BESsy.
\end{lemma}

\begin{proof}
This follows immediately from the transfer conditions of bisimilarity.
\end{proof}

The following lemma states that any structure graph obtained from a formula and an equation system is BESsy.

\begin{lemma}
For any formula $f$ and equation system $\mc{E}$, the structure graph $\tctx{\mc{E}}{f}$ is BESsy.
\end{lemma}

\begin{proof}
We have to establish that the structure graph $\tctx{\mc{E}}{f}$ is BESsy. Thereto it has to be shown that the four items of the definition of BESsyness are satisfied.

The first one trivially follows by considering all the possibilities for generating a vertex labelled by either $\top$, $\perp$, or $\fv{X}$. In each case it turns out that $f$ is of a form that does not allow the derivation of a $\rightarrow$-transition.

The proof of the second item requires induction on the depth of the proof of $\tctx{\mc{E}}{f}\up$, $\tctx{\mc{E}}{f}\down$, or $\tctx{\mc{E}}{f}\pitchfork$, respectively. Inside this induction there is a case distinction on the deduction rule that has been applied last in the proof.

For the proof of the third item it suffices to consider all possibilities for generating multiple successors and it follows easily that in these cases the vertex is also labelled by $\up$ or $\down$.

The last item follows trivially from the observation that a cycle of successor relations can never be generated without using a bound variable along the cycle. This would inevitably introduce a rank for that vertex.
\end{proof}

For a BESsy structure graph $\mc{G} = \langle T,t,\to,d,r,\nearrow \rangle$ the function
$\varphi$ is defined as follows: for $u \in T$

\[
\begin{array}{lcl}
\sgform{u} &=&
\begin{cases}
\bigsqcap \{ \sgform{u'} \mid u \rightarrow u' \} &
\mbox{if $d(u) = \up$ and $u \not\in\mathrm{dom}(r)$}, \\
\bigsqcup \{ \sgform{u'} \mid u \rightarrow u' \} &
\mbox{if $d(u) = \down$ and $u \not\in\mathrm{dom}(r)$}, \\
\true & \mbox{if $d(u) = \top$}, \\
\false & \mbox{if $d(u)=\perp$}, \\
X & \mbox{if $\nearrow(u) = X$}, \\
X_{u} & \mbox{otherwise}.
\end{cases}
\end{array}
\]

The function $\varphi$ introduces variables for those vertices that are
in the domain of the vertex rank mapping or the free variable mapping. In
the second case, the associated variable name is used. In the former case,
a fresh variable name is introduced to represent the vertex. For other
vertices the structure that is offered via vertex decoration mapping $d$
is used to obtain a formula representing such a structure.

\begin{definition}
\label{transSGtoBES}
Let $\mc{G} = \langle T,t,\to,d,r,\nearrow \rangle$ be a BESsy structure graph.
The equation system associated to $\mc{G}$, denoted $\sgbes{\mc{G}}$, is defined below.

To each vertex $u \in T$ such that $u \in \mathrm{dom}(r)$,
we associate an equation of the form
\[ \sigma X_u = \rhs{u} \]
Here
$\sigma$ is $\mu$ in case the rank associated to the vertex is odd,
and $\nu$ otherwise. $\rhs{u}$ is defined as follows:
\[
\rhs{u} = \begin{cases}
            \bigsqcap \{ \sgform{u'} \mid u \rightarrow u'\} & \text{if $d(u) = \up$} \\
            \bigsqcup \{ \sgform{u'} \mid u \rightarrow u'\} & \text{if $d(u) = \down$} \\
            \sgform{u'} & \text{if $d(u) \neq \up, d(u) \neq \down$, and $u \to u'$} \\
          \end{cases}
\]

The equation system $\sgbes{\mc{G}}$ is obtained by ordering
the equations from left-to-right ensuring the ranks of the vertices associated
to the equations are descending.
\end{definition}

We next show the correspondence between a BES and the BES obtained from its structure graph.
First, given a formula $f$ and a BES $\mc{E}$, we inductively define the set of relevant proposition variables $\kappa_{\mc{E}}(f)$
as follows:
\begin{eqnarray*}
\kappa^{0}_{\mc{E}}(f) & = & \occ{f} \\
\kappa^{n+1}_{\mc{E}}(f) & = & \kappa^{n}_{\mc{E}}(f) \cup \bigcup\{ X \mid Y \in \kappa^{n}_{\mc{E}}(f) \land \sigma Y = g \in \mc{E} \land X \in \occ{g} \} \\
\kappa_{\mc{E}}(f) & = & \kappa^{\omega}_{\mc{E}}(f)
\end{eqnarray*}
The set of relevant proposition variables contains exactly the
variables on which $f$, interpreted in the context of $\mc{E}$ depends
in some way. As long as it is clear from the context, we abbreviate
$\kappa_{\mc{E}}(f)$ to $\kappa$.

Using such a set $\kappa$ of relevant variables, we can define the BES $\mc{E}$ \emph{restricted to $\kappa$}, denoted $\mc{E}_{\kappa}$,
inductively as follows:
\begin{eqnarray*}
\epsilon_{\kappa} & = & \epsilon \\
((\sigma X = f)\mc{E})_{\kappa} & = & \begin{cases}
                                    (\sigma X = f)\mc{E}_{\kappa} & \text{if $X \in \kappa$} \\
                                    \mc{E}_{\kappa} & \text{otherwise}
                                  \end{cases}
\end{eqnarray*}

One can show that the number of
equations in $\mc{E}_{\kappa}$ is the same as the number of equations in $\sgbes{\tctx{\mc{E}}{f}}$.

More specifically, a solution preserving ordering of the
equations in $\sgbes{\tctx{\mc{E}}{f}}$ can be found such that
each equation $\sigma Y = f \in \mc{E}_{\kappa}$ corresponds to the
equation $\sigma X_{\tctx{\mc{E}}{Y}} = \rhs{\tctx{\mc{E}}{Y}} \in
\sgbes{\tctx{\mc{E}}{f}}$. Assume that $\mc{E}_{\kappa} \equiv (\sigma_1
X_1 = f_1) \dots (\sigma_n X_n = f_n)$, then $\sgbes{\tctx{\mc{E}}{f}}
\equiv (\sigma_1 X_{\tctx{\mc{E}}{X_1}} = \rhs{\tctx{\mc{E}}{X_1}})
\dots (\sigma_n X_{\tctx{\mc{E}}{X_n}} = \rhs{\tctx{\mc{E}}{X_n}})$.
Observe that in these equation systems, it suffices to show that the
right hand sides match in order to find that both equation systems have
the same solution.  The fact that the right hand sides indeed match is
shown by the following proposition.

\begin{proposition}
\label{proposition:correspondence_formula_rhs}
Let $\mc{E}$ be a BES such that $\sigma Y = f \in \mc{E}$. Then for all
environments $\eta$ for which $\eta(Z) = \eta(X_{\tctx{\mc{E}}{Z}})$ for
all $Z \in \bnd{\mc{E}}$, we have
$\sem{f}{\eta} = \sem{\rhs{\tctx{\mc{E}}{Y}}}{\eta}$.
\end{proposition}
\begin{proof}
We prove this using a distinction on the cases of $\rhs{\tctx{\mc{E}}{Y}}$.
The proof involves a number of lemmata expressing distribution laws of $\sgformname{}$ over
Boolean connectives $\land$ and $\lor$, as well as the relation between
$f$ and $\sgform{\tctx{\mc{E}}{f}}$ for arbitrary formulae $f$. These
lemmata in turn require proofs involving case distinctions on
the SOS rules, and induction on formulae.\footnote{For reviewing purposes, the required lemmata, as well as a detailed proof of this proposition (as Proposition~\ref{proposition:correspondence_formula_rhs_detailed}) can be found in the appendix.}
\end{proof}

We can combine these results to find that evaluating a formula $f$ in a BES $\mc{E}$,
and evaluating the formula $\sgform{\tctx{\mc{E}}{f}}$ in the BES $\sgbes{\tctx{\mc{E}}{f}}$ are equivalent.

\begin{theorem}\label{thm:transformation_preserves_solution}
Let $\mc{E}$ be a BES and $\eta$ an environment. Then for all formulae $f$ it holds that
$\sem{f}{}\sem{\mc{E}}{\eta} = \sem{\sgform{\tctx{\mc{E}}{f}}}{}\sem{\sgbes{\tctx{\mc{E}}{f}}}{\eta}$
\end{theorem}
\begin{proof}
The sketch of the proof is as follows. First we restrict $\mc{E}$ to the equations that are
relevant for $f$, \ie let $\kappa = \kappa_{\mc{E}}(f)$, then $\mc{E}_{\kappa}$ and
$\sgbes{\tctx{\mc{E}}{f}}$ have the same fixpoint alternations, and the equation systems can
be aligned such that each equation $\sigma Y = f \in \mc{E}_{\kappa}$ is at the same position
as the equation $\sigma X_{\tctx{\mc{E}}{Y}} = \rhs{\tctx{\mc{E}}{Y}} \in \sgbes{\tctx{\mc{E}}{f}}$.
Note that this reordering does not influence the solution.
Furthermore, using Proposition~\ref{proposition:correspondence_formula_rhs} we find that the
right-hand sides of all these equations coincide. Then also the bound variables in both BESses
have the same solution, and hence our claim follows.\footnote{For reviewing purposes a more detailed version of the proof is included as Theorem~\ref{thm:transformation_preserves_solution_detailed} in the appendix.}
\end{proof}

\section{Normalisation of Structure Graphs}
\label{Sect:normalisation}

In BESsy structure graphs, a vertex that is
decorated by a rank typically represents a proposition variable that occurs
at the left-hand side of some equation in the associated equation system,
whereas the non-ranked vertices can occur as subterms in right-hand sides of
equations with mixed occurrences of $\wedge$ and $\vee$. Normalisation
of a structure graph assigns ranks to each
non-ranked vertex that has successors. The net effect of this operation is
that the structure graph obtained thusly induces an
equation system in simple form. In choosing the rank, one has some
degree of freedom; an effective and sound strategy is to ensure that
all equations in the associated equation system end up in the very
last block. This is typically achieved by assigning $0$ as a rank.

$$
\sosrule[\label{sos:norm_and}]{t \up}{\normalise{t} \up}
\qquad
\sosrule[\label{sos:norm_or}]{t \down}{\normalise{t} \down}
\qquad
\sosrule[\label{sos:norm_edge}]{t \rightarrow t'}{\normalise{t} \rightarrow \normalise{t'}}
$$
$$
\sosrule[\label{sos:norm_true}]{t \top}{\normalise{t} \top}
\qquad
\sosrule[\label{sos:norm_false}]{t \perp}{\normalise{t} \perp}
\qquad
\sosrule[\label{sos:norm_fv}]{t \fv{X}}{\normalise{t} \fv{X}}
$$
$$
\sosrule[\label{sos:norm_ranked}]{t \pitchfork n}{\normalise{t} \pitchfork n}
\qquad
\sosrule[\label{sos:norm_unranked}]{t \not \pitchfork \quad t \to t'}
        {\normalise{t} \pitchfork 0}
$$

The last deduction rule expresses that in case a vertex $t$ does not
have a rank, rank $0$ is associated to the normalised version of
$t$, provided, of course, that the vertex has a successor. Observe that
normalisation preserves BESsyness of the structure graph, \ie, any BESsy
structure graph that is normalised again yields a BESsy structure graph.

\begin{property}
\label{property:normalised_bessy_sg}
Let $t$ be an arbitrary BESsy structure graph.
\begin{enumerate}

\item $\sgform{\normalise{t}} \in \mc{X} \cup \{\true,\false\}$;

\item $\sgbes{\normalise{t}}$ is in simple form;

\item $\normalise{\normalise{t}} \bisim \normalise{t}$.

\end{enumerate}

\end{property}

The lemmata below formalise that the solution to an equation system that
is induced by a BESsy structure graph, is preserved and reflected
by the equation system associated to the normalised counterpart of that
structure graph.

\begin{lemma}
\label{lem:normalisation1}
Let $t$ be a BESsy structure graph.
Then, there is a total injective mapping $h : \bnd{\sgbes{t}} \to
\bnd{\sgbes{\normalise{t}}}$, such that for all $\eta$:
\[\forall X \in \bnd{\sgbes{t}}:
\sem{\sgbes{t}}{\eta}(X)
=
\sem{\sgbes{\normalise{t}}}{\eta}(h(X))
\]

\end{lemma}

\begin{proof} Observe that for each ranked vertex $u$ in $t$, vertex
$\normalise{u}$ has the same rank in $\normalise{t}$. Following
Definition~\ref{transSGtoBES}, these
vertices both induce equations in the equation systems that appear in the same
block of identical fixed point equations. All unranked vertices $u'$ in
$t$ that are ranked in $\normalise{t}$, induce $\nu$-equations at the
end of the equation system induced by $\normalise{t}$. References to
these latter equations can be eliminated,
following~\cite[Lemma~6.3]{Mad:97}.
\end{proof}

\begin{lemma}
\label{lem:normalisation2}
Let $t$ be a BESsy structure graph.
Then, for all $\eta$:
\[\sem{\sgform{t}}{\sem{\sgbes{t}}{\eta}} =
\sem{\sgform{\normalise{t}}}{\sem{\sgbes{\normalise{t}}}{\eta}}\]
\end{lemma}
\begin{proof}
Follows from Lemma~\ref{lem:normalisation1}.
\end{proof}
The example below illustrates an application of normalisation, and
it provides a demonstration of the above lemmata and its implications.
\begin{example}
The BESsy structure graph depicted at the left contains a single vertex
that is not decorated with a rank. Normalisation of this structure graph
yields the structure graph depicted at the right. Assuming that
vertex $t$ is the root, $\sgbes{t}$ is as follows:
\[
\begin{array}{l}
(\mu X_u = (X_u \wedge (X_w \wedge X_w)) \vee (X_v \vee X_v))\\
(\nu X_w = (X_u \wedge (X_w \wedge X_w)) \vee (X_x \vee X_x))\\
(\mu X_v = X_v)\\
(\mu X_x = X_v \vee (X_x \vee X_x))
\end{array}
\]
$\sgbes{\normalise{t}}$ has similar top-level logical operands
as $\sgbes{t}$, but contains an extra greatest fixed point
equation trailing the other four, and references to this equation:
\[
\begin{array}{l}
(\mu X_{\normalise{u}} = X_{\normalise{t}} \vee (X_{\normalise{v}} \vee X_{\normalise{v}}))\\
(\nu X_{\normalise{w}} = X_{\normalise{t}} \vee (X_{\normalise{x}} \vee X_{\normalise{x}}))\\
(\mu X_{\normalise{v}} = X_{\normalise{v}})\\
(\mu X_{\normalise{x}} = X_{\normalise{v}} \vee (X_{\normalise{x}} \vee X_{\normalise{x}}))\\
(\nu X_{\normalise{t}} = X_{\normalise{u}} \wedge (X_{\normalise{u}} \wedge X_{\normalise{u}} ))
\end{array}
\]
\begin{center}
\parbox{.35\textwidth}
{
\begin{tikzpicture}[->,>=stealth',node distance=75pt]
\tikzstyle{every state}=[shape=rectangle,draw=none,minimum width=30pt, minimum height=20pt, inner sep=2pt]

\node[state] (X) {$u\ \text{\footnotesize{$\down\ 3$}}$};
\node[state] (Z) [below of=X] {$v\ \text{\footnotesize{$1$}}$};
\node[state] (XY)[left of=X] {$t\ \text{\footnotesize{$\up$}}$};
\node[state] (Y) [below of=XY] {$w\ \text{\footnotesize{$\down\ 2$}}$};
\node[state] (W) [below of=Y,xshift=37.5pt,yshift=35pt] {$x\ \text{\footnotesize{$\down\ 1$}}$};

\draw (X) edge  (Z) edge[bend right] (XY)
      (Y) edge (W) edge [bend right] (XY)
      (Z) edge [loop below] (Z)
      (XY) edge [bend right] (Y) edge [bend right] (X)
      (W) edge (Z) edge [loop below] (W);

\end{tikzpicture}
}
$\Longrightarrow\qquad$
\parbox{.43\textwidth}
{
\begin{tikzpicture}[->,>=stealth',node distance=75pt]
\tikzstyle{every state}=[shape=rectangle,draw=none,minimum width=30pt, minimum height=20pt, inner sep=2pt]

\node[state] (X) {$\normalise{u}\ \text{\footnotesize{$\down\ 3$}}$};
\node[state] (Z) [below of=X] {$\normalise{v}\ \text{\footnotesize{$1$}}$};
\node[state] (XY) [left of=X] {$\normalise{t}\ \text{\footnotesize{$\up\ 0$}}$};
\node[state] (Y) [below of=XY] {$\normalise{w}\ \text{\footnotesize{$\down\ 2$}}$};
\node[state] (W) [below of=Y,xshift=37.5pt, yshift=35pt] {$\normalise{x}\ \text{\footnotesize{$\down\ 1$}}$};

\draw (X) edge  (Z) edge[bend right] (XY)
      (Y) edge (W) edge [bend right] (XY)
      (Z) edge [loop below] (Z)
      (XY) edge [bend right] (Y) edge [bend right] (X)
      (W) edge (Z) edge [loop below] (W);

\end{tikzpicture}
}
\end{center}
According to Lemma~\ref{lem:normalisation1}, there is an injection
$h : \bnd{\sgbes{t}} \to \bnd{\sgbes{\normalise{t}}}$, such
that for all $X \in \bnd{\sgbes{t}}$, we have
$\sem{\sgbes{t}}{}(X) =
\sem{\sgbes{\normalise{t}}}{}(h(X))$;
$h(X_z) = X_{\normalise{z}}$ for $z \in \{u,v,w,x\}$ is such an
injection.
Following Lemma~\ref{lem:normalisation2}, we furthermore find
$\sem{\sgform{t}}{\sem{\sgbes{t}}{}} =
\sem{X_u \wedge X_w}{\sem{\sgbes{t}}{}} =
\sem{X_{\normalise{t}}}{\sem{\sgbes{\normalise{t}}}{}} =
\sem{\sgform{\normalise{t}}}{\sem{\sgbes{\normalise{t}}}{}}$.

\end{example}

The below theorem states that bisimilarity on structure graphs is a
congruence for normalisation. Ultimately, this means that the simple
form is beneficial from a bisimulation perspective: normalisation leads
to smaller quotients of structure graphs.  This addresses the hitherto open
question concerning the effect of normalisation on the bisimulation reductions
of~\cite{KeirenWillemse2009a}.

\begin{theorem}
\label{th:norm_preserv_bisimilarity}
Let $t,t'$ be arbitrary, but bisimilar structure graphs. Then also
$\normalise{t} \bisim \normalise{t'}$.
\end{theorem}

\begin{proof}
Let $R$ be a bisimulation relation witnessing $t \bisim t'$. We define
the relation
$R_n$ as $\{ (\normalise{u},\normalise{u'}) ~|~ (u,u') \in R\}$. Then
$R_n$ is a bisimulation relation witnessing $\normalise{t} \bisim
\normalise{t'}$.
\end{proof}

Finally, we show that normalisation is in fact sometimes beneficial
for bisimilarity.
\begin{example}
Consider the labelled transition system $L$ given below.
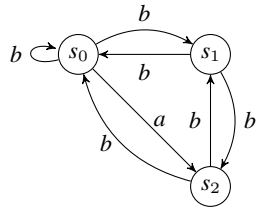
\begin{center}
\begin{tikzpicture}[->,>=stealth',node distance=50pt]
\tikzstyle{every state}=[minimum size=15pt, inner sep=2pt, shape=circle]

\node [state] (r) {\small $s_0$};
\node [state] (s) [right of=r] {\small $s_1$};
\node [state] (l) [below of=s] {\small $s_2$};

\draw
  (r) edge[loop left] node            {\small $b$} (r)
  (r) edge[bend left] node [above]    {\small $b$} (s)
  (s) edge[] node [below]             {\small $b$} (r)
  (s) edge[bend left] node [right]    {\small $b$} (l)
  (l) edge[] node [left]              {\small $b$} (s)
  (r) edge[] node [right]             {\small $a$} (l)
  (l) edge[bend left] node [left]     {\small $b$} (r);
\end{tikzpicture}
\end{center}

Let $\phi = \nu X. [ a ] X \land \langle b \rangle X$.
Consider the equation system $\E{L}{\phi}$ given below, together
with its associated structure graph:
\begin{center}
\[
\begin{array}{lcl}
(\nu X_{s_0} & = & (X_{s_2} \land X_{s_2}) \land (X_{s_0} \lor (X_{s_1} \lor X_{s_1})))\\
(\nu X_{s_1} & = & \true \land (X_{s_0} \lor (X_{s_2} \lor X_{s_2})))\\
(\nu X_{s_2} & = & \true \land (X_{s_0} \lor (X_{s_1} \lor X_{s_1})))
\end{array}
\]
\begin{tikzpicture}[->,>=stealth',node distance=50pt]
\tikzstyle{every state}=[shape=rectangle,draw=none,minimum width=30pt, minimum height=20pt, inner sep=2pt]

\node[state] (X0) {$\tctx{\mc{E}}{X_{s_0}}\ \text{\footnotesize{$\up\ 0$}}$};
\node[state] (X0X2X2) [above of=X0] {$\tctx{\mc{E}}{X_{s_0} \lor (X_{s_2} \lor X_{s_2})}\ \text{\footnotesize{$\down$}}$};
\node[state] (X0X1X1) [below of=X0] {$\tctx{\mc{E}}{X_{s_0} \lor (X_{s_1} \lor X_{s_1})}\ \text{\footnotesize{$\down$}}$};
\node[state] (X1) [left of=X0X1X1,xshift=-60pt] {$\tctx{\mc{E}}{X_{s_1}}\ \text{\footnotesize{$\up\ 0$}}$};
\node[state] (X2) [right of=X0X1X1,xshift=60pt] {$\tctx{\mc{E}}{X_{s_2}}\ \text{\footnotesize{$\up\ 0$}}$};
\node[state] (T) [below of=X0X1X1] {$\tctx{\mc{E}}{\true}\ \text{\footnotesize{$\top$}}$};

\draw (X0) edge (X2)
      (X0) edge[bend left] (X0X1X1)
      (X0X1X1) edge[bend left] (X0)
      (X0X1X1) edge (X1)
      (X0X2X2) edge (X0)
      (X0X2X2) edge (X2)
      (X1) edge (T)
      (X1) edge (X0X2X2)
      (X2) edge (T)
      (X2) edge (X0X1X1);

%
\end{tikzpicture}

\end{center}

Observe that the above structure graph can be minimised with respect to
bisimilarity, identifying vertices $\tctx{\mc{E}}{X_{s_1}}$ and
$\tctx{\mc{E}}{X_{s_2}}$, as well as $\tctx{\mc{E}}{X_{s_0} \lor (X_{s_1} \lor X_{s_1})}$ and
$\tctx{\mc{E}}{X_{s_0} \lor (X_{s_2} \lor X_{s_2})}$.
Normalising the above structure graph adds the label $\pitchfork 0$ to
vertices $\tctx{\mc{E}}{X_{s_0} \lor (X_{s_1} \lor X_{s_1})}$ and
$\tctx{\mc{E}}{X_{s_0} \lor (X_{s_2} \lor X_{s_2})}$, leading to the minimised normalised
structure graph below:
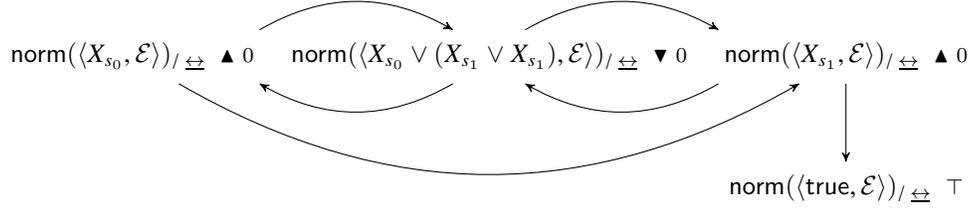
\begin{center}
\begin{tikzpicture}[->,>=stealth',node distance=50pt]
\tikzstyle{every state}=[shape=rectangle,draw=none,minimum width=30pt, minimum height=20pt, inner sep=2pt]

\node[state] (X0X1X1) {$\normalise{\tctx{\mc{E}}{X_{s_0} \lor (X_{s_1} \lor X_{s_1})}}_{/\bisim}\ \text{\footnotesize{$\down$\ 0}}$};
\node[state] (X0) [left of=X0X1X1,xshift=-85pt]{$\normalise{\tctx{\mc{E}}{X_{s_0}}}_{/\bisim}\ \text{\footnotesize{$\up\ 0$}}$};
\node[state] (X1) [right of=X0X1X1,xshift=85pt] {$\normalise{\tctx{\mc{E}}{X_{s_1}}}_{/\bisim}\ \text{\footnotesize{$\up\ 0$}}$};
\node[state] (T) [below of=X1] {$\normalise{\tctx{\mc{E}}{\true}}_{/\bisim}\ \text{\footnotesize{$\top$}}$};

\path (X1.south west) edge[bend left] (X0X1X1);
\path (X0) edge[bend right] (X1);
\path (X0.north east) edge[bend left] (X0X1X1);
\path (X0X1X1) edge[bend left] (X0.south east);
\path (X0X1X1) edge[bend left] (X1.north west);
\path (X1) edge (T);

%
\end{tikzpicture}
%
%
%
\end{center}
The above structure graph induces the following equation system:
\[
\begin{array}{l}
(\nu X_{\normalise{\tctx{\mc{E}}{X_{s_0}}}_{/\bisim}} =
X_{\normalise{\tctx{\mc{E}}{X_{s_0} \lor (X_{s_1} \lor X_{s_1})}}_{/\bisim}}
\land
(X_{\normalise{\tctx{\mc{E}}{X_{s_1}}}_{/\bisim}} \land
X_{\normalise{\tctx{\mc{E}}{X_{s_1}}}_{/\bisim}})) \\
(\nu X_{\normalise{\tctx{\mc{E}}{X_{s_1}}}_{/\bisim}} =
X_{\normalise{\tctx{\mc{E}}{X_{s_0} \lor (X_{s_1} \lor X_{s_1})}}_{/\bisim}}
\land
(\true \land \true))\\
(\nu X_{\normalise{\tctx{\mc{E}}{X_{s_0} \lor (X_{s_1} \lor X_{s_1})}}_{/\bisim}}=
X_{\normalise{\tctx{\mc{E}}{X_{s_0}}}_{/\bisim}} \lor
(X_{\normalise{\tctx{\mc{E}}{X_{s_1}}}_{/\bisim}} \lor
X_{\normalise{\tctx{\mc{E}}{X_{s_1}}}_{/\bisim}}))
\end{array}
\]
%
%
The size of $\mc{E}$ is 26. By comparison, the size of the equation
system induced by the minimised normalised structure graph is 18; one can
easily check to see that the equation system induced by the non-normalised
minimised structure graph is larger than 18. Hence this example illustrates
that $|\E{L}{\phi}_{\bisim}| > |\srf{\E{L}{\phi}}_{\bisim}|$, showing
that transforming to SRF may be beneficial to the minimising capabilities
of bisimulation.
\end{example}

\section{Bisimilarity Implies Solution Equivalence}
\label{sec:bisimilarity_vs_solution}

In this section we state one of our main results, proving that equation systems
corresponding to bisimilar BESsy structure graphs essentially have the
same solution. This allows one to safely use bisimulation minimisation
of the structure graph, and solve the equation system induced by the
minimal structure graph instead. Before we give our main theorem, we
first lift some known results for equation systems,
see \eg~\cite{Mad:97,Kei:06,KeirenWillemse2009a}, to structure graphs.\\

\begin{definition} Let $\langle T,t,\to,d,r,\fv{} \rangle$ be a structure
graph. A partial function $\choice{} {:} T \mapsto T$ is a
\emph{$\bullet$-choice function}, with $\bullet \in \{ \up,\down\}$, when
both:

\begin{itemize}
\item $\dom{\choice{}} = \{u \in T ~|~
d(u) = \bullet \wedge u \to \}$;
\item $u \to \choice{}(u)$ for all $u \in \dom{\choice{}}$.
\end{itemize}

\end{definition}
Given a $\bullet$-choice function $\choice{}$, with $\bullet
\in \{\up,\down\}$, for a structure graph, we can obtain a new structure
graph by choosing one successor among the successors for vertices
decorated with a $\bullet$, \viz, the one prescribed by $\choice{}$.
This is formalised next.

\begin{definition} Let $\mc{G} =\langle T, t, \to, d, r, \fv{} \rangle$
be an arbitrary structure graph.  Let $\bullet \in \{ \up, \down \}$,
and $\choice{}$ a $\bullet$-choice function. The structure graph
$\mc{G}_{\choice{}}$, obtained by applying the $\bullet$-choice
function $\choice{}$ on $\mc{G}$, is defined as the six-tuple
$\langle T, t, \to_{\choice{}}, d_{\choice{}}, r, \fv{}
\rangle$, where:

\begin{itemize}
\item for all $u \notin \dom{\choice{}}$,
$u \to_{\choice{}} u'$ iff $u \to u'$;

\item for all $u \in \dom{\choice{}}$, only
$u \to_{\choice{}} \choice{}(u)$;

\item $d_{\choice{}}(t) = d(t)$ and $\dom{d_{\choice{}}}  =
\{u ~|~ d(u) \not= \bullet\}$

\end{itemize}
\end{definition}
Observe that a structure graph obtained by applying a $\up$-choice
function entails a structure graph in which no vertex is labelled with
$\up$. Similarly, applying a $\down$-choice function yields
a structure graph without $\down$ labelled vertices.
\begin{property} Let $t$ be an arbitrary BESsy structure graph. Assume
an arbitrary $\bullet$-choice function $\gamma$ on $t$. Then
$\normalise{t}_{\choice{}}$ is again BESsy.

\end{property}

The effect that
applying, \eg, a $\up$-choice function has on the solution to the equation
system associated to the structure graph to which it is applied, is
characterised by the proposition below. This result is well-known in
the setting of equation systems, see, \eg~\cite{Mad:97}.

\begin{proposition}
\label{prop:conj_choice}
Let $t$ be a normalised, BESsy structure graph, with no vertex labelled
$\fv{}$.
\begin{enumerate}
 \item For all $\up$-choice functions $\choice{}$ applied to $t$,
 we have $\sem{\sgbes{t}}{} \sqsubseteq
 \sem{\sgbes{t_{\choice{}}}}{}$;

 \item There exists a $\up$-choice function $\choice{}$, such that
 $\sem{\sgbes{t}}{} = \sem{\sgbes{t_{\choice{}}}}{}$.

 \item For all $\down$-choice functions $\choice{}$ applied to $t$,
 we have $\sem{\sgbes{t}}{} \sqsupseteq
 \sem{\sgbes{t_{\choice{}}}}{}$;

 \item There exists a $\down$-choice function $\choice{}$, such that
 $\sem{\sgbes{t}}{} =
 \sem{\sgbes{t_{\choice{}}}}{}$.

\end{enumerate}
\end{proposition}
\begin{proof}
Follows immediately from \cite[Proposition 3.36]{Mad:97}, and the correspondence
between structure graphs an Boolean Equation Systems.
\end{proof}
%
%
In some cases, \viz, when a structure graph is void of any vertices
labelled $\down$ or void of vertices labelled $\up$, the solution of an
equation system associated to a structure graph can be characterised
by the structure of the graph. While one could consider these to be
degenerate cases, they are essential in our proof of the main theorem
in this section.  A key concept used in characterising the solution of
equation systems in the degenerate cases is that of a $\nu$-dominated
lasso, and its dual, $\mu$-dominated lasso.

\begin{definition}
Let $t$ be a BESsy structure graph. A lasso starting in $t$ is a finite
sequence
$t_0$, $t_1$, $\dots$, $t_n$, satisfying
$t_0= t$, $t_n = t_j$
for some $j \leq n$, and for each $1 \leq i \leq n$, $t_{i-1} \to t_{i}$.
A lasso is said to be $\nu$-dominated if $\max\{r(t_i) \mid j \leq i \leq n \}$ is even;
otherwise it is $\mu$-dominated.
\end{definition}

The following lemma is loosely based on lemmata taken from Kein\"anen
(see Lemmata 40 and 41 in \cite{Kei:06}).
\begin{lemma}
\label{lem:lassoes}
Let $t$ be a normalised, BESsy structure graph in which no vertex is
labelled with $\fv{}$. Then:

\begin{enumerate}
  \item if no vertex in $t$ is labelled with $\up$ then
        $\sem{\sgform{t}}{\sem{\sgbes{t}}{}} = \true$ iff some lasso
        starting in $t$ is $\nu$-dominated, or some maximal, finite path
        starting in $t$ terminates in a vertex labelled with $\top$;

  \item if no vertex in $t$ is labelled with $\down$ then
        $\sem{\sgform{t}}{\sem{\sgbes{t}}{}} = \false$ iff some lasso
        starting in $t$ is $\mu$-dominated, or some maximal, finite path
        starting in $t$ terminates in a vertex labelled with $\perp$

\end{enumerate}
\end{lemma}
\begin{proof}
We only consider the first statement; the proof of the second statement
is dual. Observe that since no vertex in $t$ is labelled with $\up$,
$\sgform{u} \neq \bigsqcap\{ u_1, \dots, u_n \}$ for all
$u$. We distinguish two cases:
\begin{enumerate}

\item Assume there is a $\nu$-dominated lasso $t_0, t_1, \ldots, t_n$,
starting in $t$.  BESsyness of $t$ implies that there is a ranked vertex
$t_i$ on the cycle of the lasso. Without loss of generality assume that
$t_i$ has the highest rank on the cycle of the $\nu$-dominated lasso.
By definition, this highest rank is even. This means that it
induces an equation $\nu X_{t_i} = g_i$ in $\sgbes{t}$, that precedes
all other equations $\sigma X_{t_k} = g_k$ induced by the other vertices
on the cycle. Consider the path snippet starting in $t_i$, leading
to $t_i$ again: $t_i, t_{i+1}, \ldots, t_{n-1}, t_j, t_{j+1}, t_{i-1}$.
\emph{Gauss elimination} \cite{Mad:97} allows one to substitute $g_{i+1}$
for $X_{t_{i+1}}$ in the equation for $X_{t_i}$, yielding $\nu X_{t_i}
= g_i[X_{t_{i+1}} := g_{i+1}]$. Repeatedly applying Gau\ss\ elimination
on the path snippet ultimately allows one to rewrite $\nu X_{t_i} = g_i$
to $\nu X_{t_i} = g_i' \lor X_{t_i}$, since $X_{t_{i-1}}$ depends on
$X_{t_i}$ again. The solution to $\nu X_{t_i} =  g_i' \lor X_{t_i}$ is
easily seen to be $X_{t_i} = \true$. This solution ultimately propagates
through the entire lasso, and back to $t$, leading to $\sgform{t} =
X_t = \true$.

\item Suppose there is a finite path $t_0, t_1, \ldots, t_n$ starting in
$t$, where $t_n$ is labelled with $\top$.
This means that there is an equation $\sigma X_{t_n} = \true$ on which $X_t$
depends. As the equation $\sigma X_{t_n} = \true$ is solved, we may immediately
substitute the solution in all other formulae on the path. As none of
the formulae are conjunctive, we find $\sgform{t} = \true$.

\end{enumerate}
Conversely, observe that due to Proposition~\ref{prop:conj_choice}, there is
a structure graph $t_\down$, void of any vertices labelled $\down$, that has an
equation system associated to it with solution equivalent to that of the
equation system associated to $t$. This means that $t_\down$ has no branching
structure, but is necessarily a set of lassoes and maximal, finite paths.
In case the root of $t$ is on a lasso,
$\sem{\sgform{t}}{\sem{\sgbes{t}}{}} = \true$ is because the cycle on the
lasso has an
even highest rank. In the other case, $\sem{\sgform{t}}{\sem{\sgbes{t}}{}}
= \true$ can only be the case because ultimately $t_\down$ leads to a vertex
labelled $\true$.
%
\end{proof}

Using the structure graph characterisation of solution, we prove that
for BESsy structure graphs that do not have vertices labelled with $\fv{}$,
and in which all vertices not labelled with $\top$ or $\bot$ have a rank,
bisimulation minimisation of the structure graph preserves the solution of the
associated BES.
\begin{lemma}
\label{lem:solution_closed_norm}
Let $t$, $t'$ be normalised BESsy structure graphs in which no vertex is
labelled with $\fv{}$. Assume
$t$ is minimal w.r.t strong bisimilarity.  Then $t
\bisim t'$ implies $\sem{\sgform{t}}{\sem{\sgbes{t}}{}} =
\sem{\sgform{t'}}{\sem{\sgbes{t'}}{}}$.

\end{lemma}
\begin{proof}
The case where the root of $t$ is decorated with a $\top$ or $\bot$
is trivial and therefore omitted. Assume that the root of $t$ is not
decorated with $\top$ nor $\bot$.  By Proposition~\ref{prop:conj_choice}
we know that there is a $\down$-choice function $\choice{}$
such that $\sem{\sgbes{t_{\choice{}}}}{} = \sem{\sgbes{t}}$. We next
construct a $\down$-choice function $\choice{}'$ for $t'$ that satisfies the
following condition:
\[
\forall u \in \dom{\choice{}}, u' \in \dom{\choice{}'}:~
u \bisim u' \implies \choice{}(u) \bisim \choice{}'(u)
\]
Observe that we have $t_{\choice{}} \bisim t_{\choice{}'}$, as
the choice for successors chosen in previously bisimilar $\down$-labelled vertices
are synchronised by the $\down$-choice function. Because of this bisimilarity
and the finiteness of $t'$,
any $\nu$-dominated lasso starting in a vertex $u$
reachable in $t$ implies the existence of a similar $\nu$-dominated lasso starting
in vertices $u'$ reachable in $t'$ that are bisimilar to $u$, and, of course,
also \emph{vice versa}. Likewise
for maximal finite paths. Suppose the root vertex of $t_{\choice{}}$ has only
$\nu$-dominated lassoes and finite maximal paths ending in $\top$-labelled
vertices. Then
so has $t'_{\choice{}'}$. This means that

\[
\sem{\sgform{t}}{\sem{\sgbes{t}}{}} =
\sem{\sgform{t_{\choice{}}}}{\sem{\sgbes{t_{\choice{}}}}{}} =^{\dagger}
\true =
\sem{\sgform{t'_{\choice{}'}}}{\sem{\sgbes{t'_{\choice{}'}}}{}}
{\implies}^{\!\!\!*~}
\sem{\sgform{t'}}{\sem{\sgbes{t'}}{}}
\]
At $^\dagger$, we used Lemma~\ref{lem:lassoes} and
at $^*$, we used Proposition~\ref{prop:conj_choice} to conclude
that the equation system associated to $t'_{\choice{}'}$ has a
smaller solution than the one associated to $t'$.  The case where
$\sem{\sgform{t}}{}\sem{\sgbes{t}}{} = \false$ follows the same line
of reasoning, constructing a structure graph with a $\up$-choice
function $\choice{}$, resulting in a structure graph containing no
vertices labelled $\up$.
\end{proof}

We set out to prove that bisimilar structure graphs $t$ and $t'$ always
give rise to equation systems and formulae with the same truth value. The
above lemma may seem like a roundabout way in proving this property.
In particular, the assumption in Lemma~\ref{lem:solution_closed_norm}
that $t$ is minimal with respect to bisimilarity may seem odd. The reason
for using the quotient is due to our appeal to the non-constructive
Proposition~\ref{prop:conj_choice}, as we illustrate through the
following example.
\begin{example} Consider the two bisimilar BESsy structure graphs $t$ and
$t'$ below:
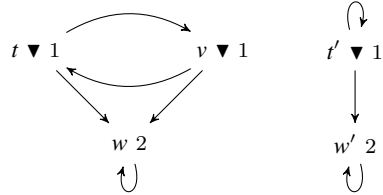
\begin{center}
\begin{tikzpicture}[->,>=stealth',node distance=50pt]
\tikzstyle{every state}=[draw=none,minimum size=15pt, inner sep=2pt,
shape=rectangle]

\node [state] (X)                    {\small $t$ \footnotesize $\down\ 1$};
\node [state] (Z) [below right of=X] {\small $w$ \footnotesize $2$};
\node [state] (Y) [above right of=Z] {\small $v$ \footnotesize $\down\ 1$};

\node [state] (X') [right of=Y] {\small $t'$ \footnotesize $\down\ 1$};
\node [state] (Z') at (X' |- Z) {\small $w'$ \footnotesize $2$};

\draw
   (X) edge [bend left] (Y) edge (Z)
   (Y) edge [bend left] (X) edge (Z)
   (Z) edge [loop below] (Z)
   (X') edge (Z') edge [loop above] (X')
   (Z') edge[loop below] (Z')
;
\end{tikzpicture}
\end{center}
Following Lemma~\ref{lem:lassoes}, we know that all vertices will be
associated to proposition variables with solution $\true$, as both
structure graphs are normalised and contain no $\up$-labelled vertices.
Appealing to Proposition~\ref{prop:conj_choice}, we know that there is
a structure graph $t_\down$ that gives rise to an equation system with
the same solution as the one that can be associated to $t$.  In fact,
there are three choices for $t_\down$:
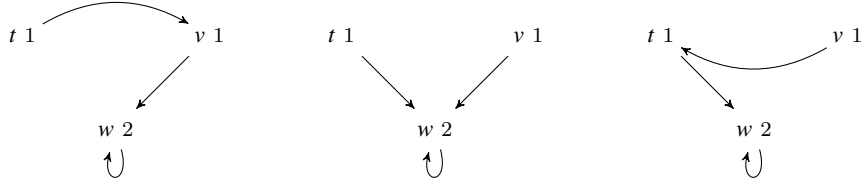
\begin{center}
\begin{tikzpicture}[->,>=stealth',node distance=50pt]
\tikzstyle{every state}=[draw=none,minimum size=15pt, inner sep=2pt,
shape=rectangle]

\node [state] (X1)                     {\small $t$ \footnotesize $1$};
\node [state] (Z1) [below right of=X1] {\small $w$ \footnotesize $2$};
\node [state] (Y1) [above right of=Z1] {\small $v$ \footnotesize $1$};
\node [state] (X2) [right of=Y1]       {\small $t$ \footnotesize $1$};
\node [state] (Z2) [below right of=X2] {\small $w$ \footnotesize $2$};
\node [state] (Y2) [above right of=Z2] {\small $v$ \footnotesize $1$};
\node [state] (X3) [right of=Y2]       {\small $t$ \footnotesize $1$};
\node [state] (Z3) [below right of=X3] {\small $w$ \footnotesize $2$};
\node [state] (Y3) [above right of=Z3] {\small $v$ \footnotesize $1$};

\draw
   (X1) edge [bend left] (Y1)
   (Y1) edge (Z1)
   (Z1) edge [loop below] (Z1)
   (X2) edge (Z2)
   (Y2) edge (Z2)
   (Z2) edge [loop below] (Z2)
   (X3) edge (Z3)
   (Y3) edge [bend left] (X3)
   (Z3) edge [loop below] (Z3)
;
\end{tikzpicture}
\end{center}
Note that all three structure graphs are associated to equation systems
with the same solution as the equation system for $t$. However, while
the middle structure graph would allow us to construct a $\down$-choice
function that resolves the choice for successors for vertex $t'$,
the other two structure graphs do not allow us to do
so, simply because they have bisimilar vertices whose only successor
leads to different equivalence classes.  Such conflicts do not arise when
assuming that $t$ is already minimal, in which case each vertex represents
a unique class.
\end{example}

Regardless of the above example, we can still derive the desired result.
Based on the previous lemma, the fact that bisimilarity is an
equivalence relation on structure graphs and the fact that quotienting
is well-behaved, we find the following theorem, which holds for arbitrary
BESsy structure graphs.

\begin{theorem}
Let $t, t'$ be arbitrary bisimilar BESsy structure
graphs. Then for all environments
$\eta$, $\sem{\sgform{t}}{}\sem{\sgbes{t}}{\eta} =
\sem{\sgform{t'}}{}\sem{\sgbes{t'}}{\eta}$.

\end{theorem}

\begin{proof}
Let $\eta$ be an arbitrary environment. Let $\overline{t}$ and
$\overline{t}'$ be the structure graphs obtained from $t$ and $t'$
by replacing all
decorations of the form $\fv{X}$ of all vertices with $\top$ if $\eta(X) = \true$,
and $\bot$ otherwise. Note that we have $\overline{t} \bisim
\overline{t}'$. Based on Lemma~\ref{lem:substitution} and
Definition~\ref{transSGtoBES}, we find:
\[
\sem{\sgform{t}}{\sem{\sgbes{t}}{\eta}}
=
\sem{\sgform{\overline{t}}}{\sem{\sgbes{\overline{t}}}{}}
\]
Likewise, we can derive such an equivalence for $\overline{t}'$ and $t'$.
By Lemma~\ref{lem:normalisation2}, we find:
\[
\sem{\sgform{\overline{t}}}{\sem{\sgbes{\overline{t}}}{}}
=
\sem{\sgform{\normalise{\overline{t}}}}
    {\sem{\sgbes{\normalise{\overline{t}}}}{}}
\]
Again, a similar equivalence can be derived for $\overline{t}'$
and $\normalise{\overline{t}'}$.  Observe that by
Theorem~\ref{th:norm_preserv_bisimilarity}, we find that $\overline{t}
\bisim \overline{t}'$ implies $\normalise{\overline{t}} \bisim
\normalise{\overline{t}'}$.  Observe that
$\normalise{\overline{t}} \bisim \quot{\normalise{\overline{t}}}
\bisim \normalise{\overline{t}'}$.
Finally, since all three are still BESsy
structure graphs, that furthermore do not contain vertices labelled with
$\fv{}$, we can apply Lemma~\ref{lem:solution_closed_norm} twice
to find:
\[
\begin{array}{ll}
& \sem{\sgform{\normalise{\overline{t}}}}
    {\sem{\sgbes{\normalise{\overline{t}}}}{}}\\
= &
\sem{\sgform{\quot{\normalise{\overline{t}}}}}
    {\sem{\sgbes{\quot{\normalise{\overline{t}}}}}{}}\\
= &
\sem{\sgform{\normalise{\overline{t}'}}}
    {\sem{\sgbes{\normalise{\overline{t}'}}}{}}
\end{array}
\]
But this necessitates our desired conclusion:
\[
\sem{\sgform{t}}
    {\sem{\sgbes{t}}{}}
=
\sem{\sgform{t'}}
    {\sem{\sgbes{t'}}{}}
\]
\end{proof}

\section{Bisimilarity on Processes vs Bisimilarity on Structure Graphs}
\label{Sect:relation}

The $\mu$-calculus and bisimilarity of labelled transition systems are
intimately related: two states in a transition system are bisimilar iff
the states satisfy the same set of $\mu$-calculus formulae. As a result,
one can rely on bisimulation minimisation techniques for reducing the
complexity of the labelled transition system, prior to analysing whether
a given $\mu$-calculus formula holds for that system. Unfortunately, in
practice, bisimulation reductions are often disappointing, and have to
be combined with safe abstractions in order to be worthwhile.

We show that minimising an equation system that encodes a model checking
problem is, size-wise, always at least as effective as first applying
a safe abstraction to the labelled transition system, subsequently minimising
the latter and only then encoding the model checking problem in an
equation system.  An additional example illustrates that bisimulation
minimisation for equation systems can in fact be more effective.

\begin{lemma}
\label{lem:basis_reflection}
 Assume $L = \langle S, \act, \to \rangle$ is an
arbitrary labelled transition system. Let $\phi$ be an arbitrary formula.
Then, for arbitrary equation system $\mc{E}$, we have:
\[
\begin{array}{rl}
\text{if} &
\forall s,s' \in S: s \bisim s' \implies
\forall \tilde X \in \bnd{\phi} \cup \occ{\phi}: \tctx{\mc{E}}{X_s}
\bisim \tctx{\mc{E}}{X_s'} \\
\text{then}
&\forall s,s' \in S: s \bisim s' \implies
\tctx{\mc{E}}{\RHS{s}{\phi}} \bisim
\tctx{\mc{E}}{\RHS{s'}{\phi}}
\end{array}
\]
\end{lemma}

\begin{proof} Assume a given equation system $\mc{E}$.
We proceed by means of an induction on the structure of $\phi$.
\begin{itemize}
\item \emph{Base cases.} Assume that for all $s,s' \in S$,
satisfying $s \bisim s'$, and all
$\tilde X \in \bnd{\phi} \cup \occ{\phi}$, we have
$\tctx{\mc{E}}{X_{s}} \bisim \tctx{\mc{E}}{X_{s'}}$.
Assume that $t,t'\in S$ are arbitrary states satisfying $t \bisim t'$.

\begin{itemize}
\item ad $\phi \equiv b$, where $b \in \{\true,\false\}$.
Clearly, $\tctx{\mc{E}}{\RHS{t}{\phi}}
= \tctx{\mc{E}}{b} = \tctx{\mc{E}}{\RHS{t'}{\phi}}$, so
bisimilarity is guaranteed by unicity of the term, regardless of the
states $t$ and $t'$;

\item ad $\phi \equiv \tilde X$. Clearly, $\tilde X \in \occ{\phi}$, so,
the required
conclusion follows immediately from the fact that
$\tctx{\mc{E}}{\RHS{t}{\phi}} =
\tctx{\mc{E}}{X_{t}} \bisim
\tctx{\mc{E}}{X_{t'}} =
\tctx{\mc{E}}{\RHS{t'}{\phi}}$;
\end{itemize}

\item Inductive cases: we assume the following induction hypothesis:
\[
\tag{IH}
\begin{array}{rl}
\text{if} & \forall s,s' \in S: s \bisim s' \implies
 \forall \tilde X \in \bnd{f_i} \cup \occ{f_i}:
\tctx{\mc{E}}{X_s} \bisim \tctx{\mc{E}}{X_{s'}} \\
\text{then} &
\forall s,s' \in S: s \bisim s' \implies
\tctx{\mc{E}}{\RHS{s}{f_i}} \bisim
\tctx{\mc{E}}{\RHS{s'}{f_i}}
\end{array}
\]
From hereon, assume that we have a pair of bisimilar states
$t,t' \in S$.

\begin{itemize}
\item ad $\phi \equiv f_1 \wedge f_2$. Assume that for any
pair of bisimilar states $s,s' \in S$, and for all
$\tilde X \in \bnd{f_1 \wedge f_2} \cup \occ{f_1 \wedge f_2} =
(\bnd{f_1} \cup \occ{f_1}) \cup (\bnd{f_2} \cup \occ{f_2})$, we have
$\tctx{\mc{E}}{X_s} \bisim
\tctx{\mc{E}}{X_{s'}}$.
By our induction hypothesis, we have $\tctx{\mc{E}}{\RHS{t}{f_1}} \bisim
\tctx{\mc{E}}{\RHS{t'}{f_1}}$ and
$\tctx{\mc{E}}{\RHS{t}{f_2}} \bisim
\tctx{\mc{E}}{\RHS{t'}{f_2}}$.
Lemma~\ref{congr} immediately
leads to $\tctx{\mc{E}}{\RHS{t}{f_1} \wedge \RHS{t}{f_2}} \bisim
\tctx{\mc{E}}{\RHS{t'}{f_1} \wedge \RHS{t'}{f_2}}$. By
definition of $\RHSname$, we have the required
$\tctx{\mc{E}}{\RHS{t}{f_1 \wedge f_2}} \bisim
\tctx{\mc{E}}{\RHS{t'}{f_1 \wedge f_2}}$.

\item ad $\phi \equiv f_1 \vee f_2$. Follows the same line of
reasoning as the previous case.

\item ad $\phi \equiv [A]f_1$.  Assume that for all pairs of
bisimilar states $s,s' \in S$, and all $\tilde X \in \bnd{[A]f_1} \cup
\occ{[A]f_1} = \bnd{f_1} \cup \occ{f_1}$, we have $\tctx{\mc{E}}{X_s}
\bisim \tctx{\mc{E}}{X_{s'}}$.  By induction, we find that
$\tctx{\mc{E}}{\RHS{s}{f_1}} \bisim \tctx{\mc{E}}{\RHS{s'}{f_1}}$ holds
for all pairs of bisimilar states $s,s'\in S$. This includes states $t$
and $t'$. Since $t$ and $t'$ are bisimilar, we have $t \xrightarrow{a}$
iff $t' \xrightarrow{a}$ for all $a \in A$.  We distinguish two cases:

\begin{enumerate}
\item Case $t \not \xrightarrow{a}$ for any $a \in A$. Then also $t'
\not \xrightarrow{a}$ for any $a \in A$.  Hence, $\RHS{t}{[A]f_1}
= \true = \RHS{t'}{[A]f_1}$.  We thus immediately have the required
$\tctx{\mc{E}}{\RHS{t}{[A]f_1}} \bisim
 \tctx{\mc{E}}{\RHS{t'}{[A]f_1}}$;

\item Case $t \xrightarrow{a}$ for some $a \in A$. Assume that $t
\xrightarrow{a} u$. Since $t \bisim t'$, we have $t' \xrightarrow{a}
u'$ for some $u' \in S$ satisfying $u \bisim u'$ (and \emph{vice
versa}). Because of our induction hypothesis, we then also have
$\tctx{\mc{E}}{\RHS{u}{f_1}} \bisim \tctx{\mc{E}}{\RHS{u'}{f_1}}$
(and \emph{vice versa}).  We thus find that for every term
in the non-empty set $\{\tctx{\mc{E}}{\RHS{u}{f_1}} \rangle ~|~ a
\in A, t \xrightarrow{a} u \}$, we can find a bisimilar term in the set
$\{\tctx{\mc{E}}{\RHS{u'}{f_1}} ~|~ a \in A, t' \xrightarrow{a} u'
\}$ and \emph{vice versa}.  Then, by Corollary~\ref{cor:capcup_props},
also $\tctx{\mc{E}}{\bigsqcap \{\RHS{u}{f_1} ~|~a \in A, t
\xrightarrow{a} u\}} \bisim \tctx{\mc{E}}{\bigsqcap \{\RHS{u'}{f_1}
~|~a \in A, t' \xrightarrow{a} u'\}}$.  This leads to
$\tctx{\mc{E}}{\RHS{t}{[A]f_1}} \bisim \tctx{\mc{E}}{\RHS{t'}{[A]f_1}}$.

\end{enumerate}
Clearly, both cases lead to the required conclusion.

\item ad $\phi \equiv \langle A \rangle f_1$. Follows the same line of
reasoning as the previous case.

\item ad $\phi \equiv \sigma \tilde X.~f_1$. Since
$\tilde X \in \bnd{\phi}$, this case follows immediately from the assumption
on $\tilde X$ and the definition of $\RHSname$.
\end{itemize}

\end{itemize}
\end{proof}

The above lemma is at the basis of the following proposition:
\begin{proposition}
\label{prop:reflection}
Let $L = \langle S, \act, \to \rangle$ be a
labelled transition system.
Let $\phi$ be an arbitrary closed $\mu$-calculus formula.
Let $s, s' \in S$ be an arbitrary pair of bisimilar states. We then have:
\[
\forall \tilde X \in \bnd{\phi}:
\langle X_{s}, \E{L}{\phi} \rangle
\bisim
\langle X_{s'}, \E{L}{\phi} \rangle
\]
\end{proposition}

\begin{proof} Let $\phi$ be an arbitrary closed formula, \ie, $\occ{\phi}
\subseteq \bnd{\phi}$; since $\phi$ is a closed formula, $\E{L}{\phi}$
will be a closed equation system. In case $\bnd{\phi} = \emptyset$,
the statement holds vacuously.
Assume $\bnd{\phi} = \{\tilde X^1,\ldots,\tilde X^n\}$, for some
$n \ge 1$. Clearly, for each variable $\tilde X^i \in \bnd{\phi}$,
we obtain equations of the form $\sigma_i X^i_s = \RHS{s}{f^i}$ in
$\E{L}{\phi}$.
Let $I$ be the relation on vertices, defined as follows:
\[
I = \{(\tctx{\E{L}{\phi}}{X^i_s}, \tctx{\E{L}{\phi}}{X^i_{s'}}
) ~|~ s,s' \in S, \tilde X^i \in \bnd{\phi}, s \bisim
s' \}
\]
According to Lemma~\ref{lem:basis_reflection}, $I$
underlies the bisimilarity between
$\tctx{\E{L}{\phi}}{\RHS{s}{f^i}}$ and
$\tctx{\E{L}{\phi}}{\RHS{s'}{f^i}}$ for pairs of bisimilar states
$s,s' \in S$. Assume $R_{f^i}$ is the bisimulation relation underlying said
equivalence. Let $R$ be defined as follows:
\[
R = I \cup \bigcup_{f^i} R_{f^i}
\]
$R$ is again a bisimulation relation, as can be checked using the
SOS rules for equations and Lemma~\ref{lem:basis_reflection}.
Clearly, $R$ relates
$\tctx{\E{L}{\phi}}{X_s}$ and $\tctx{\E{L}{\phi}}{X_{s'}}$ for arbitrary
$\tilde X \in \bnd{\phi}$ and bisimilar states $s,s' \in S$.
\end{proof}

As a result of the above proposition one can argue that bisimulation on
processes is less powerful compared to bisimulation on equation systems.
However, one may be inclined to believe that combined with abstraction,
bisimilarity on processes can lead to greater reductions. Below, we show
that even in the presence of \emph{safe} abstractions, bisimilarity on
equation systems still surpasses bisimilarity on processes.\\

We first formalise the notion of safe abstraction for processes.  Assume
$\tau$ is a constant, not present in any set of actions $\act$.
\begin{definition} An
\emph{abstraction} of a labelled transition system $L = \langle S,
\act, \to \rangle$ with respect to a set of actions $A \subseteq \act$,
is the labelled transition system $L_A = \langle S,  \act \cup \{\tau\},
\to_A \rangle$, where:

\begin{itemize}

\item for all actions $a \notin A$, $s \xrightarrow{a}_A s'$ iff
$s \xrightarrow{a} s'$;

\item $s \xrightarrow{\tau}_A s'$ iff $s \xrightarrow{a} s'$ for
some $a \in A$;

\end{itemize}
\end{definition}
In effect, an abstraction relabels an action that decorates a transition
to $\tau$ only if that action appears in the set $A$. Clearly,
if $s \bisim s'$ holds in $L$, then also $s \bisim s'$ in $L_A$, but
the converse does not hold necessarily.

\begin{definition} An abstraction $L_A$ of $L$ is said to be \emph{safe}
with respect to a closed modal $\mu$-calculus formula $\phi$ iff for each
subformula $[A']\psi$ and $\langle A' \rangle \psi$ of $\phi$, $A' \cap A =
\emptyset$.

\end{definition}
It follows from the semantics of the modal $\mu$-calculus
that all actions of some $L$, disjoint with the actions found inside the
modalities in $\phi$ can be renamed to $\tau$ without affecting the
validity of the model checking problem.
\begin{proposition} Let $L = \langle S,\act, \to \rangle$ be a
labelled transition system. Let $\phi$ be
a closed modal $\mu$-calculus formula, and assume $L_A$ is a safe abstraction of
$L$. Then for each state $s \in S$, we have $L,s \models \phi$ iff
$L_A, s \models \phi$.

\end{proposition}
The below theorem strengthens the result we obtained in
Proposition~\ref{prop:reflection}, by stating that even in the presence
of safe abstractions, bisimilarity for equation systems are as powerful
as bisimilarity taking abstractions into account.

\begin{theorem} Let $L = \langle S, \act, \to \rangle$ be an arbitrary
labelled transition system. Let $\phi$ be an arbitrary closed modal
$\mu$-calculus formula over $\act$. Then for every safe abstraction $L_A$
of $L$, we have for every pair of bisimilar states $s,s' \in S$ in $L_A$:
\[
\forall X \in \bnd{\phi}: \tctx{\E{L}{\phi}}{X_s}
\bisim \tctx{\E{L}{\phi}}{X_{s'}}
\]

\end{theorem}
\begin{proof} The proof is similar to the proof of
Proposition~\ref{prop:reflection}. In particular, it relies on
the definition of a safe abstraction to ensure that $\tctx{\mc{E}}
{\RHS{s}{[A']\psi}}$ and $\tctx{\mc{E}}{\RHS{s'}{[A']\psi}}$ for states
$s,s'$ that are bisimilar in $L_A$, but not in $L$, are mapped onto
$\tctx{\mc{E}}{\true}$ for both LTSs.
\end{proof}
Lastly, we provide an example that demonstrates that bisimulation
reduction of equation systems can lead to arbitrarily larger reductions
compared to the reductions achievable through safe abstractions and
minimisation of a given LTS. This provides the ultimate proof that
bisimilarity for equation systems surpasses that for processes.

\begin{example} Let $N$ be an arbitrary positive number. Consider
the process described by the following set of recursive processes
(using process algebra style notation):
\[
\{ P_1 = a\cdot Q_N, \quad P_{n+1} = a \cdot P_n, \quad
   Q_1 = b\cdot P_N, \quad Q_{n+1} = b \cdot Q_n ~|~ n < N \}
\]
Process $P_N$ induces an LTS $L$ that performs a sequence of $a$ actions
of length $N$, followed by a sequence of $b$ actions of length $N$,
returning to process $P_N$.  Observe that the process $P_N$ cannot
be reduced further modulo bisimulation.  Let $\phi$ be the modal
$\mu$-calculus formula $\phi = \nu \tilde X.~ \langle \{a,b\} \rangle \tilde X
$, asserting that there is an infinite
sequence consisting of $a$'s, $b$'s, or $a$'s and $b$'s. Clearly,
there is no safe abstraction of process $P_N$ with respect to $\phi$,
other than process $P_N$ itself.
The equation system $\E{P_N}{\phi}$ is as follows:
\[
\begin{array}{ll}
\nu \{  (X_{P_1} = X_{Q_N} \vee X_{Q_N}),
       (X_{P_{n+1}} = X_{P_n} \vee X_{P_n}), \\
\phantom{\nu \{}
       (X_{Q_1} = X_{P_N} \vee X_{P_N}),
       (X_{Q_{n+1}} = X_{Q_n} \vee X_{Q_n})
     ~|~ n < N \}
\end{array}
\]
We find that $\tctx{\E{P_N}{\phi}}{X_{P_N}}$ and
$\tctx{(\nu Y = Y \vee Y)}{Y}$ are bisimilar, which demonstrates
a reduction of a factor $2N$. As the labelled transition system can be
scaled to arbitrary size, this demonstrates that bisimilarity for
equation systems can be arbitrarily more effective, \ie
$|\E{L_{/\bisim}}{\phi}| > |\E{L}{\phi}_{/\bisim}|$.

\end{example}

\section{Application}
\label{Sect:Application}

Equation systems that are not immediately in simple form can be obtained
through the reduction of process equivalence checking problems such as
the branching bisimulation problem, see \eg~\cite{CPPW:07}, and the more
involved model checking problems.  As a slightly more involved example of
the latter, we analyse an unreliable channel using $\mu$-calculus model
checking.  The channel can read messages from its environment through
the $r$ action, and send or lose these next through the $s$ action and
the $l$ action, respectively.  In case the message is lost, subsequent
attempts are made to send the message until this finally succeeds; this
is achieved through some internal system behaviour modelled by action
$i$. The labelled transition system, modelling this system is given below.

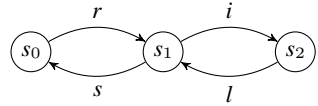
\begin{center}
\begin{tikzpicture}[->,>=stealth',node distance=50pt]
\tikzstyle{every state}=[minimum size=15pt, inner sep=2pt, shape=circle]

\node [state] (r) {\small $s_0$};
\node [state] (s) [right of=r] {\small $s_1$};
\node [state] (l) [right of=s] {\small $s_2$};

\draw
  (r) edge[bend left] node [above] {\small $r$} (s)
  (s) edge[bend left] node [below]    {\small $s$} (r)
  (s) edge[bend left] node [above]    {\small $i$} (l)
  (l) edge[bend left] node [below]    {\small $l$} (s);
\end{tikzpicture}
\end{center}

Suppose we wish to verify for which states it holds whether along all
paths consisting of reading and sending actions, it is infinitely often
possible to potentially never perform a send action. Intuitively, this
should be the case in all states: from states $s_0$ and $s_1$, there
is a finite path leading to state $s_1$, which can subsequently produce
the infinite path $(s_1\ s_2)^\omega$, along which the send action does
not occur.  For state $s_2$, we observe that there is no path consisting
of reading and sending actions, so the property holds vacuously in $s_2$.
We formalise this problem as follows:\footnote{Alternative phrasings
are possible, but this one nicely projects onto an equation system with
non-trivial right-hand sides, clearly illustrating the theory outlined
in the previous sections in an example of manageable proportions.}
\[
\phi \equiv
\nu \tilde X.~ \mu \tilde Y.~ ( ([\{r,s\}] \tilde X \wedge (\nu \tilde Z.~
\langle \overline{s} \rangle \tilde Z) )
 \vee [\{r,s\}]\tilde Y)
\]
Verifying which states in the labelled transition system satisfy $\phi$
is answered by solving the below equation system. Note that the equation
system was obtained through Definition~\ref{def:transformation}. The
solution to $X_{s_i}$ answers whether $s_i \models \phi$.
\[
\begin{array}{l}
(\nu X_{s_0} = Y_{s_0})\\
(\nu X_{s_1} = Y_{s_1})\\
(\nu X_{s_2} = Y_{s_2})\\

(\mu Y_{s_0} = ((X_{s_1} \wedge X_{s_1}) \wedge Z_{s_0}) \vee
((Y_{s_1} \wedge Y_{s_1}) \vee (Y_{s_1} \wedge Y_{s_1})) )\\
(\mu Y_{s_1} = ((X_{s_0} \wedge X_{s_0}) \wedge Z_{s_1}) \vee
((Y_{s_0} \wedge Y_{s_0}) \vee (Y_{s_0} \wedge Y_{s_0})) )\\
(\mu Y_{s_2} = (\true \wedge Z_{s_2}) \vee \true)\\

(\nu Z_{s_0} = Z_{s_1} \vee Z_{s_1})\\
(\nu Z_{s_1} = Z_{s_2} \vee Z_{s_2})\\
(\nu Z_{s_2} = Z_{s_1} \vee Z_{s_1})
\end{array}
\]
An answer to the global model checking problem would be encoded by the structure graph
$\tctx{\E{L}{\phi}}{X_{s_0} \wedge X_{s_1} \wedge X_{s_1}}$. We here only
depict the structure graph encoding the local model checking problem
$s_0 \models \phi$, encoded by the structure graph $\tctx{\E{L}{\phi}}{X_{s_0}}$,
which has root $t_1$.
Note that the ranked vertices $t_i$ originate from the $i$-th equation
in the equation system. Likewise, the unranked vertices $u_i$ originate from the
right-hand side of the $i$-th equation.
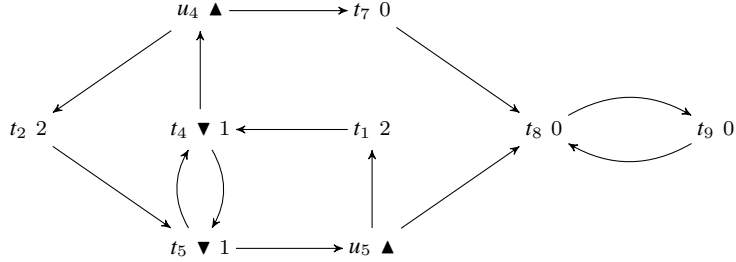
\begin{center}
\begin{tikzpicture}[->,>=stealth',node distance=65pt]
\tikzstyle{every state}=[draw=none,minimum size=15pt, inner sep=2pt, shape=rectangle]

\node [state] (Xs0)                            {\small $t_1$ \footnotesize $2$};
\node [state] (Ys0)          [left of=Xs0]     {\small $t_4$ \footnotesize $\down\ 1$};
\node [state] (Xs1Zs0)       [above of=Ys0,yshift=-20pt]    {\small $u_4$ \footnotesize $\up$};
\node [state] (Ys1)          [below of=Ys0,yshift=20pt]    {\small $t_5$ \footnotesize $\down\ 1$};
\node [state] (Xs1)          [left of=Ys0]     {\small $t_2$ \footnotesize $2$};
\node [state] (Xs0Zs1)       [right of=Ys1]    {\small $u_5$ \footnotesize $\up$};
\node [state] (Zs0)          [right of=Xs1Zs0] {\small $t_7$ \footnotesize $0$};
\node [state] (Zs1)          [right of=Xs0]    {\small $t_8$ \footnotesize $0$};
\node [state] (Zs2)          [right of=Zs1]    {\small $t_9$ \footnotesize $0$};

\draw
   (Xs0) edge (Ys0)
   (Ys0) edge (Xs1Zs0) edge[bend left] (Ys1)
   (Ys1) edge[bend left] (Ys0) edge (Xs0Zs1)
   (Xs1Zs0) edge (Xs1)
   (Xs1) edge (Ys1)
   (Xs0Zs1) edge (Xs0) edge (Zs1)
   (Xs1Zs0) edge (Zs0)
   (Zs0) edge (Zs1)
   (Zs1) edge[bend left] (Zs2)
   (Zs2) edge[bend left] (Zs1)
   ;
\end{tikzpicture}
\end{center}
Observe that we have $t_1 \bisim t_2$, $t_7 \bisim t_8 \bisim t_9$,
$t_4 \bisim t_5$ and $u_4 \bisim u_5$. Minimising the above
structure graph with respect to bisimulation leads to the structure
graph depicted below:

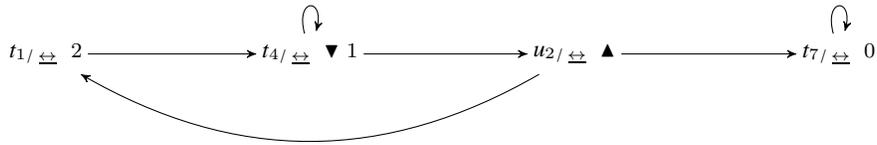
\begin{center}
\begin{tikzpicture}[->,>=stealth',node distance=100pt]
\tikzstyle{every state}=[draw=none,minimum size=15pt, inner sep=2pt, shape=rectangle]

\node [state] (Xs0_r)                            {\small $\quot{{t_1}}$ \footnotesize $2$};
\node [state] (Ys0_r)      [right of=Xs0_r]  {\small $\quot{{t_4}}$ \footnotesize $\down\ 1$};
\node [state] (Xs1Zs0_r)   [right of=Ys0_r]  {\small $\quot{{u_2}}$ \footnotesize $\up$};
\node [state] (Zs0_r)      [right of=Xs1Zs0_r] {\small $\quot{{t_7}}$ \footnotesize $0$};

\draw
   (Xs0_r) edge (Ys0_r)
   (Ys0_r) edge (Xs1Zs0_r) edge[loop above] (Ys0_r)
   (Xs1Zs0_r) edge (Zs0_r) edge[bend left] (Xs0_r)
   (Zs0_r) edge [loop above] (Zs0_r)
   ;
\end{tikzpicture}
\end{center}
Note that the structure graph is BESsy, and, hence, admits a translation
back to an equation system. Using the translation provided in
Definition~\ref{transSGtoBES} results in the following equation system:
\[
\begin{array}{l}
(\nu X_{\quot{{t_1}}} = X_{\quot{{t_4}}})\\
(\mu X_{\quot{{t_4}}} = (X_{\quot{{t_7}}} \wedge (X_{\quot{{t_1}}} \wedge X_{\quot{{t_1}}}) )
 \vee (X_{\quot{{t_4}}} \vee X_{\quot{{t_4}}}))\\
(\nu X_{\quot{{t_7}}} = X_{\quot{{t_7}}} )
\end{array}
\]
Answering the verification problem $s_0 \models \phi$ problem can thus be achieved
by solving 3 equations rather than the original 9 equations. Using
standard algorithms for solving equation systems, one quickly finds
that all equations of the minimised equation system (and thereby most of
the equations from the original equation system they represent)
have $\true$ as their solutions.
Note that the respective sizes of the equation systems are 52 before
minimisation and 14 after minimisation, which is
almost a 75\% gain; even when counting only the required equations in
the original equation system, one still has a 65\% gain.
Such gains appear to be typical in this setting (see
also~\cite{KeirenWillemse2009a}), and surpass those in the setting
of labelled transition systems. Similar gains are found for the global model
checking problem. Observe, moreover, that the original
labelled transition system already is minimal, demonstrating once more
that the minimisation of an equation system can be more effective than
minimising the original labelled transition system.

\section{Closing Remarks}
\label{Sect:Conclusions}

\paragraph*{Summary}
We presented a set of deduction rules for deriving \emph{structure
graphs} from proposition formulae and Boolean equation systems,
following the regime of~\cite{Plotkin04a}.  In defining these rules, we
focused on simplicity. We carefully selected a small set of
computationally cheap logical equivalences that we wished to be reflected
by bisimilarity in our structure graphs, and subsequently showed that
we met these goals.

Structure graphs generalise the \emph{dependency graphs} of
\eg~\cite{Mad:97,Kei:06}. The latter formalism is incapable of
capturing all the syntactic riches of Boolean equation systems, and is
only suited for a subset of closed equation systems in simple form.
A question, put forward in~\cite{KeirenWillemse2009a}, is how these
restrictions affect the power of reduction of strong bisimulation.
In Section~\ref{Sect:normalisation}, we showed that these restrictions
are in fact beneficial to the identifying power of bisimilarity. This
result follows immediately from the meta-theory for structured
operational rules, see \eg~\cite{Mousavi05-IC}.  We furthermore proved
that also in our richer setting, bisimulation minimisation of a
structure graph, induced by an equation system, preserves
and reflects the solution to the original equation system. This
generalises~\cite[Theorem~1]{KeirenWillemse2009a} for dependency graphs.

Beyond the aforementioned results, we studied the connection between
bisimilarity for labelled transition systems, the $\mu$-calculus
model checking problem and bisimilarity for structure graphs.
In Section~\ref{Sect:relation}, we showed that bisimulation minimisation
of a structure graph (associated to an equation system encoding an
arbitrary model checking problem on an arbitrary labelled transition
system) is at least as effective as bisimulation minimisation of the
labelled transition system prior to the encoding.  This relation even
holds when bisimilarity is combined with safe abstractions for labelled
transition systems.  We moreover show that this relation is strict through
an example formula $\phi$ and a labelled transition system $L$ of $2N$
($N \ge 1$) states that is already minimal (even when considering safe
abstractions with respect to $\phi$), whereas the structure graph induced
by the equation system encoding the model checking problem can be reduced
by a factor $2N$. These results provide the theoretical underpinning for the
huge reductions observed in~\cite{KeirenWillemse2009a}.

\paragraph*{Outlook}

The structure graphs that we considered in this paper are of both
theoretical and practical significance. They generalise various
graph-based models, including the aforementioned dependency graphs,
but also Parity Games~\cite{Zie:98}, and there are strong links between
our structure graphs and Switching Graphs~\cite{GP:09}. Given these
links, a \emph{game-based} characterisation of the concept of solution
for equation systems, stated in terms of our choice functions and
structure graphs is open for investigation.  In general, we consider
studying equivalences weaker than bisimilarity for structure graphs
to be worthwhile. For instance, it is not immediately clear whether the
\emph{idempotence-identifying bisimilarity} of~\cite{KeirenWillemse2009a},
which weakens some of the requirements of strong bisimilarity while
preserving and reflecting the solution of the equation system, carries
over to structure graphs without significant modifications. Furthermore,
it would be very interesting to study variations of stuttering equivalence
in this context, as it is one of the few equivalence relations that
allow for good compression at favourable computational complexities.

A thorough understanding of the structure graphs, and the associated
notions of bisimilarity defined thereon, can also be seen as a first
step towards defining similar-spirited notions in the setting of
\emph{parameterised Boolean equation systems}~\cite{GW:05b}. The latter
are high-level, symbolic descriptions of (possibly infinite) Boolean
equation systems.  The advantage of such a theory would be that it
hopefully leads to more elegant and shorter proofs of various PBES
manipulations that currently require lengthy and tedious (transfinite)
inductive proofs.

\bibliographystyle{acmtrans}
\bibliography{sosbesjournal}

\newpage
\appendix

\section{Detailed proofs and additional lemmata}

\begin{lemma}
\label{lem:form_over_and_or}
Let $f, g$ be formulae, $\mc{E}$ a BES, and $\eta$ an arbitrary environment, then we have the
following semantic equivalences:
$$
\begin{array}{rcl}
\sem{\sgform{\tctx{\mc{E}}{f}} \land \sgform{\tctx{\mc{E}}{g}}}{\eta} & = & \sem{\sgform{\tctx{\mc{E}}{f \land g}}}{\eta}\\
\sem{\sgform{\tctx{\mc{E}}{f}} \lor \sgform{\tctx{\mc{E}}{g}}}{\eta} & = & \sem{\sgform{\tctx{\mc{E}}{f \lor g}}}{\eta}
\end{array}
$$
\end{lemma}
\begin{proof}
We prove the first statement. Proof of the second statement is completely symmetric.

We first prove the implication $\sem{\sgform{\tctx{\mc{E}}{f \land g}}}{\eta} \Rightarrow \sem{\sgform{\tctx{\mc{E}}{f}} \land \sgform{\tctx{\mc{E}}{g}}}{\eta}$.
We use induction on the structure of $\sgform{\tctx{\mc{E}}{f \land g}}$:
\begin{itemize}
  \item case $\sgform{\tctx{\mc{E}}{f \land g}} = \bigsqcap\{ \sgform{u'} \mid \tctx{\mc{E}}{f \land g} \to u' \}$. It follows that $d(\tctx{\mc{E}}{f \land g}) = \up$ and $\tctx{\mc{E}}{f \land g} \not \in \dom{r}$. As $d(\tctx{\mc{E}}{f \land g}) = \up$ and $\tctx{\mc{E}}{f \land g}$ is BESsy, there must be at least one $u'$ such that $\tctx{\mc{E}}{f \land g} \to u'$.

  We need to show that for each conjunct $u' \in \{ \sgform{u'} \mid \tctx{\mc{E}}{f \land g} \to u' \}$ either $u' \in \{ \sgform{u''} \mid \tctx{\mc{E}}{f} \to u''\}$ or $u' \in \{ \sgform{u''} \mid \tctx{\mc{E}}{g} \to u''\}$, or $u' = \sgform{\tctx{\mc{E}}{f}}$, or $u' = \sgform{\tctx{\mc{E}}{g}}$. Let $u'$ be an arbitrary conjunct in $\{ \sgform{u'} \mid \tctx{\mc{E}}{f \land g} \to u' \}$. So we know $\tctx{\mc{E}}{f \land g} \to u'$. We apply case distinction on the inference rules that can introduce this edge.
  \begin{itemize}
    \item $\tctx{\mc{E}}{f \land g} \to u'$ is introduced through rule \sosref{sos:and_left_not_ranked}.
    Then we may assume that $d(\tctx{\mc{E}}{f}) = \up$, $\tctx{\mc{E}}{f} \not \in \dom{r}$ and $\tctx{\mc{E}}{f} \to u'$. According to the definition of $\sgformname$ we find that $\sgform{\tctx{\mc{E}}{f}} = \sqcap\{ \sgform{u''} \mid \tctx{\mc{E}}{f} \to u'' \}$. Hence by induction we find that $u'$ is a conjunct of $\sgform{\tctx{\mc{E}}{f}}$. As $\sgform{\tctx{\mc{E}}{f}}$, every conjunct of this formula is also a conjunct of \sgform{\tctx{\mc{E}}{f \land g}}.

    \item $\tctx{\mc{E}}{f \land g} \to u'$ is introduced through rule \sosref{sos:and_right_not_ranked}. This case is analogous to the previous case.

    \item $\tctx{\mc{E}}{f \land g} \to u'$ is introduced through rule \sosref{sos:and_not_and_left}.
    We may assume that $\neg \tctx{\mc{E}}{f} \up$. Therefore, $u' = \tctx{\mc{E}}{f}$, and the corresponding formula is $\sgform{\tctx{\mc{E}}{f}}$.

    \item The cases where $\tctx{\mc{E}}{f \land g} \to u'$ is introduced through rules \sosref{sos:and_not_and_right}, \sosref{sos:and_ranked_left} or \sosref{sos:and_ranked_right} are analogous to the previous case.
  \end{itemize}

  \item case $\sgform{\tctx{\mc{E}}{f \land g}} = \bigsqcup\{ \sgform(u') \mid \tctx{\mc{E}}{f \land g} \to u' \}$. According to rule \sosref{sos:and_ax} it must be the case that $\tctx{\mc{E}}{f \land g} \up$. According to BESsyness then $d(\tctx{\mc{E}}{f \land g}) \neq \down$, hence $\sgform{\tctx{\mc{E}}{f \land g}} \neq \bigsqcup\{ \sgform(u') \mid \tctx{\mc{E}}{f \land g} \to u' \}$, hence this case cannot apply.

  \item the cases where $\sgform{\tctx{\mc{E}}{f \land g}} \in \{ \true, \false, X \}$ are analogous to the previous case.
  \item case $\sgform{\tctx{\mc{E}}{f \land g}} = X_{\tctx{\mc{E}}{f \land g}}$. Appealing to rule \sosref{sos:and_ax} it must be the case that $\sgform{\tctx{\mc{E}}{f \land g}} \up$. Furthermore we know $\tctx{\mc{E}}{f \land g} \in \dom{r}$. According to rule \sosref{sos:rank_ax} all ranked terms are of the form
  $\tctx{\mc{E}}{Y}$, for some $Y$. This contradicts the assumption that the term we are considering is $\tctx{\mc{E}}{f \land g}$.
\end{itemize}

The reverse case, showing that $\sem{\sgform{\tctx{\mc{E}}{f \land g}}}{\eta} \Leftarrow \sem{\sgform{\tctx{\mc{E}}{f}} \land \sgform{\tctx{\mc{E}}{g}}}{\eta}$ commences by induction on the structure of $\sgform{\tctx{\mc{E}}{f}}$ and $\sgform{\tctx{\mc{E}}{g}}$. We show that each conjunct of
$\sgform{\tctx{\mc{E}}{f}}$ is also a conjunct of $\sgform{\tctx{\mc{E}}{f \land g}}$. The case for $\sgform{\tctx{\mc{E}}{g}}$ is completely analogous.
\begin{itemize}
  \item case $\sgform{\tctx{\mc{E}}{f}} = \bigsqcap\{\varphi(u') \mid \tctx{\mc{E}}{f} \to u' \}$.
  In this case we know that $d(\tctx{\mc{E}}{f}) = \up$, and $\tctx{\mc{E}}{f} \not \in \dom{r}$. Let $\tctx{\mc{E}}{f} \to u'$, so $\varphi(u')$ is a top level conjunct of $\sgform{\tctx{\mc{E}}{f}}$. From rule \sosref{sos:and_left_not_ranked} it follows immediately that $\tctx{\mc{E}}{f \land g} \to u'$, and $d(\tctx{\mc{E}}{f \land g}) = \up$ according to \sosref{sos:and_ax}, hence $\sgform{\tctx{\mc{E}}{f \land g}} = \bigsqcap\{ \sgform{u'} \mid \tctx{\mc{E}}{f \land g} \to u' \}$, and $u'$ is a conjunct of $\sgform{\tctx{\mc{E}}{f \land g}}$.

  \item $\sgform{\tctx{\mc{E}}{f}} = \bigsqcup\{\varphi(u') \mid \tctx{\mc{E}}{f} \to u' \}$. So we know that $d(\tctx{\mc{E}}{f}) = \down$ and $\tctx{\mc{E}}{f} \not \in \dom{r}$. Observe that the only conjunct of $\sgform{\tctx{\mc{E}}{f}}$ is $\sgform{\tctx{\mc{E}}{f}}$ itself. We show that $\sgform{\tctx{\mc{E}}{f}}$ is a conjunct of $\sgform{\tctx{\mc{E}}{f \land g}}$. According to rule \sosref{sos:and_not_and_left}, $\tctx{\mc{E}}{f \land g} \to \tctx{\mc{E}}{f}$. Furthermore $d(\tctx{\mc{E}}{f \land g}) = \up$ according to \sosref{sos:and_ax} and $\tctx{\mc{E}}{f \land g} \not \in \dom{r}$ according to \sosref{sos:rank_ax}, hence $\sgform{\tctx{\mc{E}}{f \land g}} = \sqcap\{ \sgform{u'} \mid \tctx{\mc{E}}{f \land g} \to u' \}$, and $\sgform{\tctx{\mc{E}}{f}}$ is a conjunct of $\sgform{\tctx{\mc{E}}{f \land g}}$.

  \item cases $\sgform{\tctx{\mc{E}}{f}} \in \{ \true, \false, X \}$ follow a similar line of reasoning as the previous case.

  \item $\sgform{\tctx{\mc{E}}{f}} = X_{\tctx{\mc{E}}{f}}$, where $\tctx{\mc{E}}{f} \in \dom{r}$. This again follows a similar line of reasoning. We use the observation that the only edge that is generated from $\tctx{\mc{E}}{f \land g}$ induced by $\tctx{\mc{E}}{f}$ is the edge $\tctx{\mc{E}}{f \land g} \to \tctx{\mc{E}}{f}$ because $f$ is ranked, according to \sosref{sos:and_ranked_left}, and in case also $d(\tctx{\mc{E}}{f}) \not \in \{\up, \down\}$ the same edge is generated (according to rule \sosref{sos:and_not_and_left}).
\end{itemize}
\end{proof}

\begin{lemma}
\label{lem:form_vs_sg_form}
Let $\mc{E}$ be a BES, $\eta$ an environment, such that $\eta(Y) = \eta(X_{\tctx{\mc{E}}{Y}})$ for all
$Y \in \bnd{\mc{E}}$. Let $f$ be a formula, such that $\occ{f} \subseteq \{ Y \mid X_{\tctx{\mc{E}}{Y}} \in \bnd{\sgbes{\tctx{\mc{E}}{f}}} \cup \free{\sgbes{\tctx{\mc{E}}{f}}} \}$. Then it holds
that $\sem{f}{\eta} = \sem{\sgform{\tctx{\mc{E}}{f}}}{\eta}$
\end{lemma}
\begin{proof}
Let $\mc{E}$ be this BES, and $f$ a formula. Assume that $\occ{f} \subseteq \{ Y \mid X_{\tctx{\mc{E}}{Y}} \in \bnd{\sgbes{\tctx{\mc{E}}{f}}} \cup \free{\sgbes{\tctx{\mc{E}}{f}}} \}$.
We show that $\sem{f}{\eta} = \sem{\sgform{\tctx{\mc{E}}{f}}}{\eta}$ by induction on the structure of $f$.

\begin{itemize}
  \item $f = \true$. By definition of $\sgformname$, $\sem{\sgform{\tctx{\mc{E}}{\true}}}{\eta} = \sem{\true}{\eta}$.
  \item $f = \false$. Analogous to the previous case.
  \item $f = Y$. We distinguish two cases, either $Y$ is bound, or $Y$ is free:
    \begin{itemize}
      \item $Y$ is bound, \ie $X_{\tctx{\mc{E}}{Y}} \in \bnd{\sgbes{\tctx{\mc{E}}{f}}}$. We derive:
      $$
      \begin{array}{ll}
        & \sem{\sgform{\tctx{\mc{E}}{Y}}}{\eta}\\
      = & \{ \text{$X_{\tctx{\mc{E}}{Y}} \in \sgbes{\tctx{\mc{E}}{f}}$, hence $\tctx{\mc{E}}{Y} \in \dom{r}$, use definition of $\sgformname$} \}\\
        & \sem{X_{\tctx{\mc{E}}{Y}}}{\eta}\\
      = & \{ \text{Semantics of BES} \}\\
        & \eta(X_{\tctx{\mc{E}}{Y}}) \\
      = & \{ \text{Assumption $\eta(X_{\tctx{\mc{E}}{Y}}) = \eta(Y)$} \}\\
        & \eta(Y)\\
      = & \{ \text{Semantics of BES} \}\\
        & \sem{Y}{\eta}
      \end{array}
      $$
      \item $Y \in \free{\sgbes{\tctx{\mc{E}}{f}}}$. This case is easy, as $Y \in \free{\sgbes{\tctx{\mc{E}}{f}}}$, also $\fv{\tctx{\mc{E}}{Y}}{Y}$, hence using the definition of $\sgformname$ we immediately find $\sem{\sgform{\tctx{\mc{E}}{Y}}}{\eta} = \sem{Y}{\eta}$.
    \end{itemize}
  \item $f = g \land g'$. Based on the SOS we know that $d(\tctx{\mc{E}}{g \land g'}) = \up$. As induction hypothesis we assume that the lemma holds for all subformulae. We derive:
  $$
  \begin{array}{ll}
    & \sem{\sgform{\tctx{\mc{E}}{g \land g'}}}{\eta}\\
  = & \{ \text{Lemma~\ref{lem:form_over_and_or}} \}\\
    & \sem{\sgform{\tctx{\mc{E}}{g}} \land \sgform{\tctx{\mc{E}}{g'}}}{\eta}\\
  = & \{ \text{Semantics of BES} \}\\
    & \sem{\sgform{\tctx{\mc{E}}{g}}}{\eta} \land \sem{\sgform{\tctx{\mc{E}}{g'}}}{\eta}\\
  = & \{ \text{Induction hypothesis} \}\\
    & \sem{g}{\eta} \land \sem{g'}{\eta}\\
  = & \{ \text{Semantics of BES} \}\\
    & \sem{g \land g'}{\eta}
  \end{array}{ll}
  $$
  \item $f = g \lor g'$. Analogous to the previous case.
\end{itemize}
\end{proof}

\begin{lemma}
\label{lem:sgform_vs_rhs}
Let $\mc{E}$ be a BES, $(\sigma X = f) \in \mc{E}$. Then it holds that
$\sgform{\tctx{\mc{E}}{f}} = \rhs{\tctx{\mc{E}}{X}}$.
\end{lemma}
\begin{proof}
Assume that $(\sigma X = f) \in \mc{E}$. Observe that $\tctx{\mc{E}}{X} \in \dom{r}$.
We show this lemma using case distinction on rules for $\rhsname$.
\begin{itemize}
  \item $d(\tctx{\mc{E}}{X}) = \up$. Then according to rule \sosref{sos:var_and} also $d(\tctx{\mc{E}}{f}) = \up$, and furthermore $\tctx{\mc{E}}{f} \not \in \dom{r}$. We derive:
  $$
  \begin{array}{ll}
  & \rhs{\tctx{\mc{E}}{X}}\\
= & \{ \text{Definition of \rhsname} \}\\
  & \bigsqcap\{\varphi(u') \mid \tctx{\mc{E}}{X} \to u' \}\\
= & \{ \text{$d(\tctx{\mc{E}}{f}) = \up$ and $\tctx{\mc{E}}{X} \not \in \dom{r}$, hence $\tctx{\mc{E}}{X} \to u'$ iff $\tctx{\mc{E}}{f} \to u'$ according to rule \sosref{sos:var_edge_and_non_ranked_rhs}} \}\\
  & \bigsqcap\{\varphi(u') \mid \tctx{\mc{E}}{f} \to u' \}\\
= & \{ \text{Definition of \sgformname} \}\\
  & \sgform{\tctx{\mc{E}}{f}}
  \end{array}
  $$
  \item $d(\tctx{\mc{E}}{X}) = \down$. Analogous to the previous case.
  \item $d(\tctx{\mc{E}}{X}) \neq \up$ and $d(\tctx{\mc{E}}{X}) \neq \down$. We know that there is exactly one $u'$ such that $\tctx{\mc{E}}{X} \to u'$, hence using rule \sosref{sos:var_edge_simple_rhs} we find $\tctx{\mc{E}}{X} \to \tctx{\mc{E}}{f}$. By definition of $\rhsname$, $\rhs{\tctx{\mc{E}}{X}} = \sgform{\tctx{\mc{E}}{f}}$.
\end{itemize}
\end{proof}

\begin{proposition}[(Proposition~\ref{proposition:correspondence_formula_rhs} in the main text)]
\label{proposition:correspondence_formula_rhs_detailed}
Let $\mc{E}$ be a BES such that $\sigma Y = f \in \mc{E}$. Then for all
environments $\eta$ for which $\eta(Y) = \eta(X_{\tctx{\mc{E}}{Y}})$,
$\sem{f}{\eta} = \sem{\rhs{\tctx{\mc{E}}{Y}}}{\eta}$.
\end{proposition}
\begin{proof}
We prove this using a distinction on the cases of $\rhs{\tctx{\mc{E}}{Y}}$.
\begin{itemize}
  \item case $d(\tctx{\mc{E}}{Y}) = \up)$.
  We derive:
    $$
    \begin{array}{ll}
      & \sem{\rhs{\tctx{\mc{E}}{Y}}}{\eta} \\
    = & \{ \text{Lemma~\ref{lem:sgform_vs_rhs}, $\sigma Y = f \in \mc{E}$} \}\\
      & \sem{\sgform{\tctx{\mc{E}}{f}}}{\eta} \\
    = & \{ \text{Lemma~\ref{lem:form_vs_sg_form}} \}\\
      & \sem{f}{\eta}
    \end{array}
    $$
  \item The cases where $d(\tctx{\mc{E}}{Y}) = \down$ and $d(\tctx{\mc{E}}{Y}) \not \in \{ \up, \down \}$ are completely analogous.
\end{itemize}
\end{proof}

\begin{lemma}
\label{lem:corresponding_rhses}
Let $\mc{E}_n \equiv (\sigma_1 X_1 = f_1) \dots (\sigma_n X_n = f_n)$,
$\mc{E}'_n \equiv (\sigma_1 X'_1 = f'_1) \dots (\sigma_n X'_n = f'_n)$.
If for all environments $\eta$ such that for all $Y$ $\eta(Y) = \eta(Y')$, it holds that for all $i$, $1 \leq i \leq n : \sem{f_i}{\eta} = \sem{f'_i}{\eta}$ then for all
$\eta'$ that satisfy for all $Z \in \occ{\mc{E}} \setminus \bnd{\mc{E}} : \eta'(Z) = \eta'(Z')$
it holds for all $X_i \in \bnd{\mc{E}}$, that $\sem{\mc{E}_n}{\eta'}(X) = \sem{\mc{E}'_n}{\eta'}(X')$.
\end{lemma}
\begin{proof}
We prove this by induction on $n$.
\begin{itemize}
  \item case $n = 0$, this case is trivial.
  \item case $n = k + 1$. Denote
  $$\begin{array}{lclcl}
  \mc{E}_{k+1} &\equiv& (\sigma_0 X_0 = f_0)\mc{E}_{k} &\equiv& (\sigma_0 X_0 = f_0)(\sigma_1 X_1 = f_1) \dots (\sigma_k X_k = f_k)\\
  \mc{E}'_{k+1} &\equiv& (\sigma_0 X'_0 = f'_0)\mc{E}'_{k} &\equiv& (\sigma_0 X'_0 = f'_0)(\sigma_1 X'_1 = f'_1) \dots (\sigma_k X'_k = f'_k)
  \end{array}$$

  Assume that for all $\Theta$ satisfying for all $Y$: $\Theta(Y) = \Theta(Y')$, it holds that for $1 \leq i \leq k$: $\sem{f_i}{\Theta} = \sem{f'_i}{\Theta}$

  Let $\Theta'$ be an arbitrary environment, such that $\forall Z \in \occ{\mc{E}_{k+1}} \setminus \bnd{\mc{E}_{k+1}}$ it holds that $\Theta'(Z) = \Theta'(Z')$.
  Let $X_i \in \bnd{\mc{E}_{k+1}}$.

  We show that
  $ \sem{ (\sigma_0 X_0 = f_0)\mc{E}_{k}}{\Theta'}(X_i) = \sem{(\sigma_0 X'_0 = f'_0)\mc{E}'_{k}}{\Theta'}(X'_i)$ for $\sigma_0 = \nu$; the case for $\sigma_0 = \mu$ is completely analogous.
  We derive the following:
  \[
  \begin{array}{ll}
     & \sem{(\sigma_0 X_0 = f_0)\mc{E}_{k}}{\Theta'}(X_i) \\
   = & \{ \text{Semantics of BES} \} \\
     & \sem{\mc{E}_{k}}{\Theta'[X_0 := \sem{f_0}{}\sem{\mc{E}_{k}}{\Theta'[X_0 := \true]}]}(X_i)\\
   = & \{ X'_0 \not \in \occ{\mc{E}_{k}} \cup \bnd{\mc{E}_{k}} \cup \occ{f_0} \} \\
     & \sem{\mc{E}_{k}}{\Theta'[X_0 := \sem{f_0}{}\sem{\mc{E}_{k}}{\Theta'[X_0, X'_0 := \true]}]}(X_i)\\
   = & \{ \text{Induction hypothesis} \} \\
     & \sem{\mc{E}_{k}}{\Theta'[X_0 := \sem{f_0}{}\sem{\mc{E}'_{k}}{\Theta'[X_0, X'_0 := \true]}]}(X_i)\\
   = & \{ \text{Assumption on $\Theta'$} \} \\
     & \sem{\mc{E}_{k}}{\Theta'[X_0 := \sem{f'_0}{}\sem{\mc{E}'_{k}}{\Theta'[X_0, X'_0 := \true]}]}(X_i)\\
   = & \{ X_0 \not \in \occ{\mc{E}'_{k}} \cup \bnd{\mc{E}'_{k}} \cup \occ{f'_0} \} \\
     & \sem{\mc{E}_{k}}{\Theta'[X_0 := \sem{f'_0}{}\sem{\mc{E}'_{k}}{\Theta'[X'_0 := \true]}]}(X_i)\\
   = & \{ X'_0 \not \in \bnd{\mc{E}_{k}} \cup \occ{\mc{E}_{k}} \} \\
     & \sem{\mc{E}_{k}}{\Theta'[X_0, X'_0 := \sem{f'_0}{}\sem{\mc{E}'_{k}}{\Theta'[X'_0 := \true]}]}(X_i)\\
   = & \{ \text{Induction hypothesis} \} \\
     & \sem{\mc{E}'_{k}}{\Theta'[X_0, X'_0 := \sem{f'_0}{}\sem{\mc{E}'_{k}}{\Theta'[X'_0 := \true]}]}(X_i)\\
   = & \{ X_0 \not \in \bnd{\mc{E}'_{k}} \cup \occ{\mc{E}'_{k}} \} \\
     & \sem{\mc{E}'_{k}}{\Theta'[X'_0 := \sem{f'_0}{}\sem{\mc{E}'_{k}}{\Theta'[X'_0 := \true]}]}(X_i)\\
   = & \{ \text{Semantics of BES} \} \\
     & \sem{(\sigma_0 X'_0 = f'_0)\mc{E}'_{k}}{\Theta'}(X'_i)
  \end{array}
  \]
\end{itemize}
\end{proof}

Recall the definition of $\kappa$, extracting the relevant variables and equations.
Given a formula $f$ and a BES $\mc{E}$, we inductively define the set of relevant proposition variables $\kappa$
as follows:
\begin{eqnarray*}
\kappa^{0}_{\mc{E}}(f) & = & \occ{f} \\
\kappa^{n+1}_{\mc{E}}(f) & = & \kappa^{n}_{\mc{E}}(f) \cup \bigcup\{ X \mid Y \in \kappa^{n}_{\mc{E}}(f) \land \sigma Y = g \in \mc{E} \land X \in \occ{g} \} \\
\kappa_{\mc{E}}(f) & = & \kappa^{\omega}_{\mc{E}}(f)
\end{eqnarray*}
The set of relevant proposition variables contains exactly the variables on which $f$, interpreted in the context of $\mc{E}$ depends in some way.

Using such a set $\kappa$ of relevant equations, we can define the BES $\mc{E}$ \emph{restricted to $\kappa$}, denoted $\mc{E}_{\kappa}$,
inductively as follows:
\begin{eqnarray*}
\epsilon_{\kappa} & = & \epsilon \\
((\sigma X = f)\mc{E})_{\kappa} & = & \begin{cases}
                                    (\sigma X = f)\mc{E}_{\kappa} & \text{if $X \in \kappa$} \\
                                    \mc{E}_{\kappa} & \text{otherwise}
                                  \end{cases}
\end{eqnarray*}

\begin{property}
Let $\mc{E}$ be a BES, and $f$ a formula. $(\sigma X = g) \in \mc{E}_{\kappa}$ implies that $\rank{\mc{E}_{\kappa}}{X} = \rank{\sgbes{\tctx{\mc{E}}{f}}}{X_{\tctx{\mc{E}}{X}}}$.
\end{property}

\begin{theorem}[(Theorem~\ref{thm:transformation_preserves_solution} in the main text)]
\label{thm:transformation_preserves_solution_detailed}
Let $\mc{E}$ be a BES and $\eta$ an environment. Then for all formulae $f$ it holds that
$\sem{f}{}\sem{\mc{E}}{\eta} = \sem{\sgform{\tctx{\mc{E}}{f}}}{}\sem{\sgbes{\tctx{\mc{E}}{f}}}{\eta}$
\end{theorem}
\begin{proof}
First we restrict $\mc{E}$ to the equations that are
relevant for $f$, \ie let $\kappa = \kappa_{\mc{E}}(f)$, than $\mc{E}_{\kappa}$ and
$\sgbes{\tctx{\mc{E}}{f}}$ have the same fixpoint alternations, and the equation systems can
be aligned such that each equation $\sigma Y = f \in \mc{E}_{\kappa}$ is at the same position
as the equation $\sigma X_{\tctx{\mc{E}}{Y}} = \rhs{\tctx{\mc{E}}{Y}} \in \sgbes{\tctx{\mc{E}}{f}}$. In
other words, we have
$\mc{E}_{\kappa} \equiv (\sigma_1 Y_1 = f_1) \dots (\sigma_n Y_n = f_n)$ and
$\sgbes{\tctx{\mc{E}}{f}} \equiv (\sigma_1 X_{\tctx{\mc{E}}{Y_1}} =  \rhs{\tctx{\mc{E}}{Y_1}}) \dots (\sigma_n X_{\tctx{\mc{E}}{Y_n}} =  \rhs{\tctx{\mc{E}}{Y_n}})$.

Observe that for all $\eta$, satisfying for all $Y$, $\eta(Y) = \eta(X_{\tctx{\mc{E}}{Y}})$, it holds that for all $i$, $1 \leq i \leq n : \sem{f_i}{\eta} = \sem{\rhs{\tctx{\mc{E}}{Y_i}}}{\eta}$ using Proposition~\ref{proposition:correspondence_formula_rhs_detailed}. Our conclusion that both solutions are equivalent now follows immediately from Lemma~\ref{lem:corresponding_rhses}.
\end{proof}

\begin{received}
\end{received}

\end{document}